\documentclass[%
 reprint,
superscriptaddress,
 amsmath,amssymb,
 aps,
]{revtex4-2}

\usepackage{graphicx}
\usepackage{dcolumn}
\usepackage{bm}

\usepackage{wrapfig}
\usepackage{amsmath}
\usepackage{booktabs}
\usepackage{amsfonts}
\usepackage{amsthm}
\usepackage{comment}
\usepackage{color,soul}
\usepackage{xcolor,colortbl}
\usepackage{wrapfig}
\usepackage{float}
\usepackage{tabularx}
\usepackage{booktabs}
\usepackage{multirow}

\definecolor{Gray}{gray}{0.85}
\definecolor{LightCyan}{rgb}{0.88,1,1}
\newcolumntype{a}{>{\columncolor{Gray}}c}
\newcolumntype{b}{>{\columncolor{white}}c}

\begin{document}

\preprint{APS/123-QED}

\newcommand{\change}[1]{\textcolor{black}{#1}}


\title{Data-driven model construction for  anisotropic dynamics of active matter}

\author{Mengyang Gu}
\email{Corresponding author: mengyang@pstat.ucsb.edu}
 \affiliation{Department of Statistics and Applied Probability, University of California, Santa Barbara, Santa Barbara, California 93106, USA}%

\author{Xinyi Fang}
\affiliation{Department of Statistics and Applied Probability, University of California, Santa Barbara, Santa Barbara, California 93106, USA}%

\author{Yimin Luo}
\email{Corresponding author: yimin.luo@yale.edu}
\affiliation{Department of Mechanical Engineering and Materials Science, Yale University, New Haven, Connecticut 06511, USA}%


\begin{abstract}
 
 The dynamics of cellular pattern formation is crucial for understanding embryonic development and tissue morphogenesis. Recent studies have shown that human dermal fibroblasts cultured on liquid crystal elastomers can exhibit an increase in orientational alignment over time, accompanied by cell proliferation, under the influence of the weak guidance of a molecularly aligned substrate. 
However, a comprehensive understanding of how this order arises remains largely unknown. This knowledge gap may be attributed, in part, to a scarcity of mechanistic models that can capture the temporal progression of the complex nonequilibrium dynamics during the cellular alignment process. The orientational alignment occurs primarily when cells reach a high density near confluence. Therefore, for accurate modeling, it is crucial to take into account both the cell-cell interaction term and the influence from the substrate, acting as a one-body external potential term.
To fill in this gap, we develop a hybrid procedure that utilizes statistical learning approaches to extend the state-of-the-art physics models {for quantifying both effects}. {We develop a more efficient way to perform feature selection that avoids testing all feature combinations through simulation.}  The maximum likelihood estimator of the model was derived and implemented in computationally scalable algorithms for model calibration and simulation. By including these features, such as the non-Gaussian, anisotropic fluctuations, and limiting alignment interaction only to neighboring cells with the same velocity direction, this model {quantitatively} reproduce the key system-level parameters: the temporal progression of the velocity orientational order parameters and the variability of velocity vectors, {whereas models missing any of the features fail to capture these temporally dependent parameters}. 
The computational tools we develop for automating model construction and calibration can be applied to other systems of active matter.

\end{abstract}
\maketitle
\setstcolor{blue}

\section{Introduction}


Active matter refers to systems composed of many individual agents that interact with each other and consume energy from their surroundings or an internal source to generate complex, global behaviors \cite{marchetti2013hydrodynamics,sanchez2012spontaneous,ramaswamy2017active}.  It encompasses a wide range of biological systems, such as flocks of birds, schools of fish, groups of humans, herds of sheep, monolayers of cells,  colonies of bacteria, and others, all exhibiting intriguing out-of-equilibrium phenomena. The dynamics of active matter, which we term ``active dynamics'', is the key to  describing the collective, spatiotemporal self-organization that arises from interactions and decision-making at the individual agent level. 

Disorder-to-order transitions in biological tissues, characterized by the emergence of patterns, are a common feature of cellular systems. 
These transitions have been compared to phases of matter, such as disordered gas and amorphous solids \cite{yang2021configurational,angelini2011glass}. They underlie many developmental processes \cite{lenne2022sculpting}, and are implicated in cancer invasion \cite{ray2018dynamics}. The resulting  patterns impart tissue with form, function, and integrity, as seen in the basket-weave-like pattern of the dermis that serves as a shield for the deeper layers 
\cite{biggs2020mechanical,ferguson2004scar}, the generation of forces in blood vessels \cite{sarkar2006development,choi2014circumferential}, and the coordination of cell fates during embryonic development
\cite{hinnant2020coordinating}. 

The mechanisms behind these cellular transitions are not well understood, but traditionally, collective behaviors are thought to be regulated by cell chemosensing \cite{camley2016emergent}, and upstream biochemical signaling pathways \cite{ridley2003cell}. While biochemical regulations are on-demand and short-lived, physical guidance ensures the generation of cell phenotypes and long-term maintenance of tissue structures \cite{tonndorf2021isotropic}. 
More recently, simple two-dimensional transitions of such nature have been realized through \textit{in vitro} experiments \cite{gregoire2003moving,turiv2020topology,babakhanova2020cell,martella2019liquid}, by subjecting a cell monolayer to an external guiding field, such as a well-aligned molecular field \cite{luo2022cell}, which can lead cells to collectively orient along a predetermined axis in a density-dependent manner. 

Modeling spontaneous alignment of particles in active matter systems has been the subject of numerous studies  \cite{vicsek1995novel,toner1998flocks,szabo2006phase,chate2008collective,ginelli2010large,li2017mechanism}. The two main approaches in modeling are either by constructing mechanistic  models through physical laws, or by applying data-driven methods to extract information from observations to construct models \cite{bongard2007automated,bruckner2021learning}. 
 {Describing individual cell behaviors often starts from a Langevin equation of Brownian particles \cite{zwanzig2001nonequilibrium}: $m\dot{v}=-\zeta v+\epsilon_F$, with $m$, $v$ and $\zeta$ being the mass, velocity and friction coefficient, respectively, and  $\epsilon_F$ being a random force fluctuation, typically assumed to be a Gaussian white noise.  Variants of it have been widely used for modeling persistent random motions of cells in experiments and simulation \cite{selmeczi2005cell,wu2014three,bruckner2019stochastic,bruckner2020inferring,ferretti2020building}, whereas the complex cell-cell interaction at high density, often implicated in tissue formation and repair, is not explicitly considered in these models.} 

In data-driven methods, regression techniques, for instance, are widely applied to estimate linear coefficients of a set of additive basis  \cite{chou2009recent,rudy2017data,reinbold2020using}, and to learn particle interaction kernel functions by piece-wise linear functions \cite{lu2019nonparametric}, neural networks \cite{colen2021machine}, and  Gaussian processes  \cite{gu2022scalable}. {Learning the distance-based interaction kernel function has been successfully applied to  model schools of golden shiner
\cite{katz2011inferring}, flocks of surf scoters \cite{lukeman2010inferring} and systems of interacting particles \cite{lu2019nonparametric}. However, these approaches are rarely applied to estimate cell-cell interaction from experiments with a large number of cellular trajectories, partly because   the large computational cost prohibits accurate interaction kernel learning approaches \cite{gu2022scalable}. Furthermore, as the underlying mechanism of these interactions remains largely unknown, a principled way to  identify the relevant input of interaction kernel and delineating neighboring sets 
is needed as adding nonrelevant inputs can dramatically reduce estimation efficiency.}  

Despite the prevalence of the disorder-to-order transition observed in experiments \cite{luo2022cell,duclos2014perfect,garcia2015physics},  capturing the temporal progression of dynamic quantities in active matter systems, such as the  orientational order parameter and variability of velocity, remains a challenging task due to a few reasons. First of all, active matter systems intrinsically contain large fluctuations. Gaussian distributions of fluctuation, for instance, are often assumed for modeling the velocity progression in constructing mechanistic models \cite{fily2012athermal,bechinger2016active}. In a data-driven  approach, the model is usually optimized by minimizing a loss function. One common choice of the loss function is the sum of squared errors, and minimizing it is equivalent to finding the maximum likelihood estimator, under the assumption of having independent Gaussian noises with equal variances. However, we find that velocity distributions from our experiments significantly deviate from Gaussian. In fact, even though models assuming Gaussian distribution of the velocity fluctuation can fit the magnitude of temporally dependent velocity, they cannot capture the progression of orientational order parameters. 
Non-Gaussian distributions of displacements have been widely observed in biological systems, such as particles absorbed on lipid bilayers,  immersed in entangled actin solutions \cite{wang2009anomalous}, in a bath of swimming algal cells \cite{leptos2009dynamics}, liposomes in active actin solutions \cite{wang2012brownian}, and receptors on living cell membranes \cite{he2016dynamic}. 
While models have been developed to explain the fluctuation of displacements independent of time \cite{dieterich2008anomalous,cherstvy2018non} {and for how these distributions arise \cite{czirok1998exponential}}, the fluctuation distributions have rarely been incorporated in simulation to reproduce the temporal progression of velocity and order parameters in nonequilibrium cell migrations.

Second, in principle, having a large number of observations containing thousands of particle trajectories over hundreds of time points can help offset the uncertainty of estimation due to  large fluctuations of the active particles. Nonetheless, calibrating the simulation model with such a large number of observations is a nontrivial task.
{Placing a smooth Bayesian prior on the interaction kernel function, such as a Gaussian process prior, can partially filter the noise in estimation but also leads to large computational costs. Thus, most of the current studies of learning interaction kernels are restricted to a small number of particles  and time frames \cite{lu2019nonparametric,miller2020learning}. } 
Hence, there is a need for robust and computationally scalable algorithms for {feature selection and estimation} from a large number of observations. 

This work aims to address these issues by introducing an efficient workflow   to automate model construction, promoting convergence between simulation and experiments. 
We take a hybrid approach that integrates physical models and statistical learning approaches: First, we start from a conventional form of the physical models (e.g. the Vicsek model \cite{vicsek1995novel,toner1995long}). Second, we apply statistical tests for feature selections. Then, these features are utilized to construct a  data-generative model, instead of minimizing a loss function as usually adopted in other data-driven discovery approaches. 
The data generative model provides a better physical interpretation of the estimated quantities, and more importantly, the uncertainty of the estimation can also be quantified.  We develop an automated feature selection and  estimation approach,  which is applied to learning physical quantities from  live microscopy of fibroblasts moving on liquid crystal elastomer substrates, to discover a variety of new features. Simulation models constructed with these new features can reproduce the progression of system properties, such as orientational order parameters and velocity distributions, whereas models missing any of these features do not match experimental findings. 

 \begin{figure*}[t] 
\centering
\includegraphics[width=\textwidth]{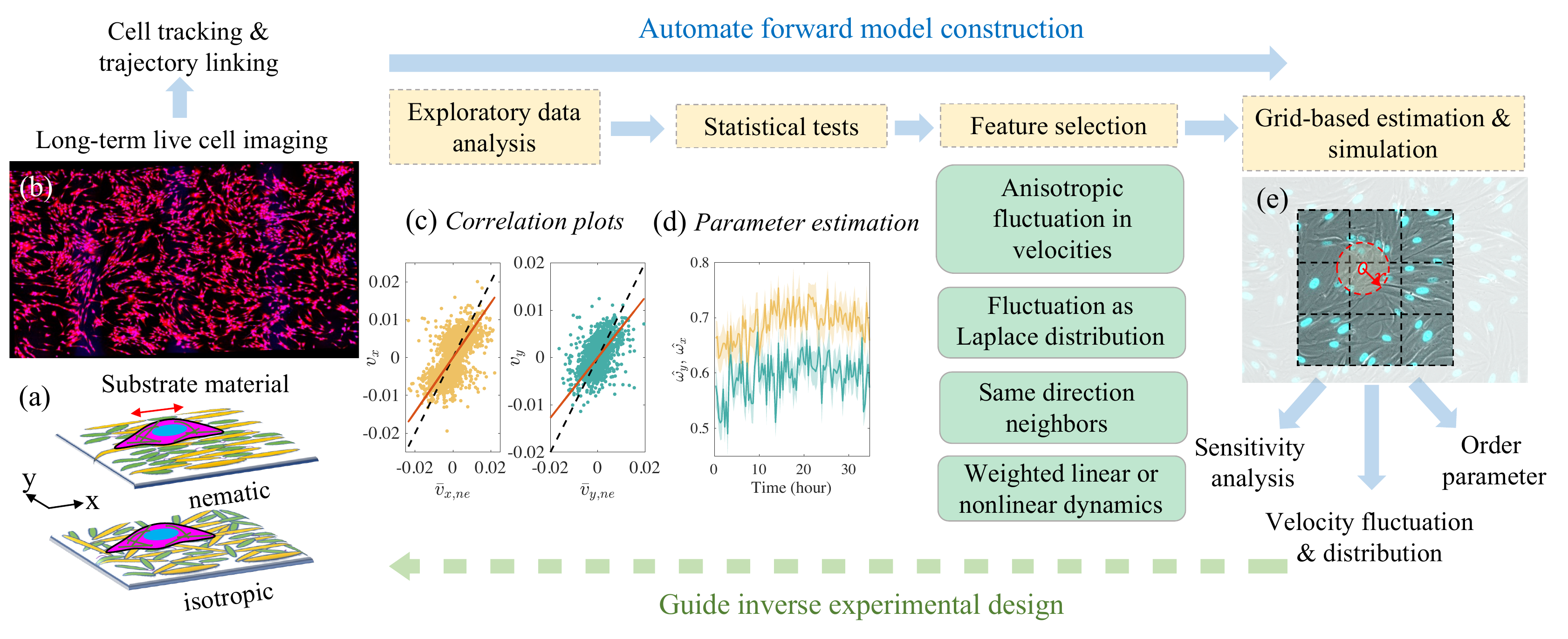}
   \caption{(a) Schematics of the two types of substrates (isotropic and nematic) tested in this study. {On the nematic substrate, the molecular alignment (denoted by a red double-sided arrow), is parallel to x.} (b) A stitched view of a snapshot of long-term cell imaging where cell cytoplasm is labeled in red. (c) Correlation plots in exploratory data analysis, where x, y velocity components are plotted against those of the neighboring  average in the previous step. The slope of the fit (red solid lines) is smaller than 1 (black dashed lines). (d) Parameter estimation of the slope parameters, $\omega$, the slope value of the red solid line in (c), and their $95\%$ confidence intervals (shaded area). The statistical tests yield four potential features (green blocks), representing new features added to the classical Vicsek model in fitting the current data set. (e)  The grid-based system to search for the neighbors. In order to find the neighbor of the target cell $i$ (circled in red), a radius of $r$ will be searched. The computational procedure is simplified by only searching through neighboring squares connected to the square cell $i$ is located in. {This workflow generates various physical parameters via simulation to compare to quantities computed from the experiment.} 
   }
\label{fig:overview}
\end{figure*}

Our main contribution  is to develop an efficient feature selection and estimation approach for cellular movement experiments on liquid crystal elastomer substrates from  video microscopy. The four {selected} features can be classified into two groups.  The first group  concerns cell-substrate interactions. We found that the velocity of the cells is anisotropic and has 
 heavier tails than a Gaussian distribution, motivating the use of a Laplace distribution or generalized Gaussian distribution of fluctuation for characterizing cell-substrate interaction. 
Though Laplace distribution was used to fit the probability distribution of the displacements    \cite{chaudhuri2007universal,wang2009anomalous,wang2012brownian} {and the mechanism of exponentially distributed cellular motility was explored in \cite{czirok1998exponential}},  its effect on capturing the progression of the orientational order parameter in simulations has not been demonstrated. 
 The second group of discoveries concern cell-cell interactions. {We observe from experiments that when cells interact, cell bodies become elongated, suggesting cells pulling each other along through a tensional network.} 
 Our findings indicate that cells are not perfectly aligned with the average of their neighbors in the prior time frame, as is the case in the classic Vicsek model \cite{vicsek1995novel}; Instead, effects from the previous step shrink towards zero, likely due to a frictional force term in the Langevin equation \cite{zwanzig2001nonequilibrium}.
  Additionally, we discovered that cells traveling in the opposite direction may be excluded from the neighboring set, leading to improved predictions of the velocity at the subsequent time point. This likely arises from the nematic nature of cell-cell interactions \cite{duclos2014perfect}, where cells {traveling in the opposite direction} can simply glide past each other without influencing each other's velocity. 
  
{Furthermore, we develop scalable computational tools for analyzing and simulating dynamical processes of active matter}. Figure \ref{fig:overview} illustrates an instance of the particle-based  module of our computational tools. 
In Fig. \ref{fig:overview}(c), we present a plot showing the correlation between cellular velocity and the mean velocity of neighboring cells in the previous time point. The slopes of the best-fit lines (red solid curve) are smaller than 1 for both $x$ and $y$ directions. To determine if this result applies to all time points, we plot the fitted weight parameters (equivalent to the slope of the fit in Fig. \ref{fig:overview}(c)) over time and calculate the 95\% confidence interval, which is represented by the shaded area in Fig. \ref{fig:overview}(d). The upper bounds of the 95\% confidence intervals are substantially smaller than 1, indicating that the weight parameters must be included in the simulation model. All the identified features are included in the simulation model, and a maximum likelihood estimator is derived for efficient parameter estimation (Appendixes A and B). For both parameter estimation and simulation, we employ a grid-based approach \cite{ginelli2016physics} by storing the particles in a coarse-grained grid (Fig. \ref{fig:overview}(e)), {which can improve the efficiency in searching for neighbors}.  
With a typical microscopy video of n $\approx$ 2500 cells and T $\approx$  100 time frames, the time it takes to perform feature selection, parameter estimation, and simulation of dynamics with particle interactions, is less than 30 seconds on a desktop computer. This provides almost immediate feedback that can be used to inversely guide the design of the experiment. The data sets and code used in the article have been made publicly available \cite{code_cell_model_2023}.

\section{Feature selection from experimental results}

We begin by discussing experimental methods for live cell imaging and the process of extracting trajectories from cells cultured on aligned (nematic) and disordered (isotropic) liquid crystal elastomer substrates. 
We then conduct exploratory data analysis and statistical tests to identify significant features from the experimental data. These features are used as building blocks to extend a baseline model. Lastly, we perform simulations to validate our findings in Section III. 

\subsection{Experimental setting: {\MakeLowercase{In vitro}} cell alignment experiment}
\label{sec:feature_selection}

The experimental data were derived from \cite{luo2022cell}, {consisting of cell trajectories.} Liquid crystal elastomer (LCE) substrates were fabricated with a mixture of reactive monomers RM82 and RM23 (SYNTHON Chemicals GmbH \& Co.) in a 1:1 molar ratio 
with the molecular field having either isotropic or nematic (uniform along the $x$-direction) configurations.  {The monomers were crosslinked to make a solid film, which was topographically flat but molecularly aligned, previously examined by wide angle X-ray scattering and atomic force microscopy \cite{luo2022cell}. }
Human dermal fibroblasts (HDFs, American Type Culture Collection) 
were seeded onto the substrates at cell density $\rho \approx$ 50 mm$^{-2}$, and grown to desired $\rho$. Prior to imaging, cell nuclei and { cytoplasm} were stained with Dyes Hoechst 33342 {and CellTracker Deep Red} (both from ThermoFisher), respectively.
Cells are maintained at 37 $^o$C, 5\% CO$_2$ 
and imaged every 20 min for about 36 hours. Tens of images were taken at every time point and stitched together to construct an image on the order of square millimeters in area.  LCE nematic substrates impose a molecular direction to guide HDF growth. During the experiment, cell density $\rho$ grows from roughly 350 to 400 mm$^{-2}$. With an increase in cell density, the monolayer undergoes a disorder-to-order transition. {Initially, there were approximately 2600 cells captured within the frame. When we finished imaging, the count had increased to about 2950 cells.} To obtain the trajectories as inputs, we fit an ellipse to the nuclei and track particle trajectories using  ImageJ \cite{schneider2012nih}, a standard  cellular imaging tool.
On the nematic substrate, cells develop a long-range order along the $x$-direction, the same direction as the molecular orientation in the aligned substrate, but not on isotropic substrates. 

 \begin{figure*} 
\centering
\includegraphics[width=0.85\textwidth]{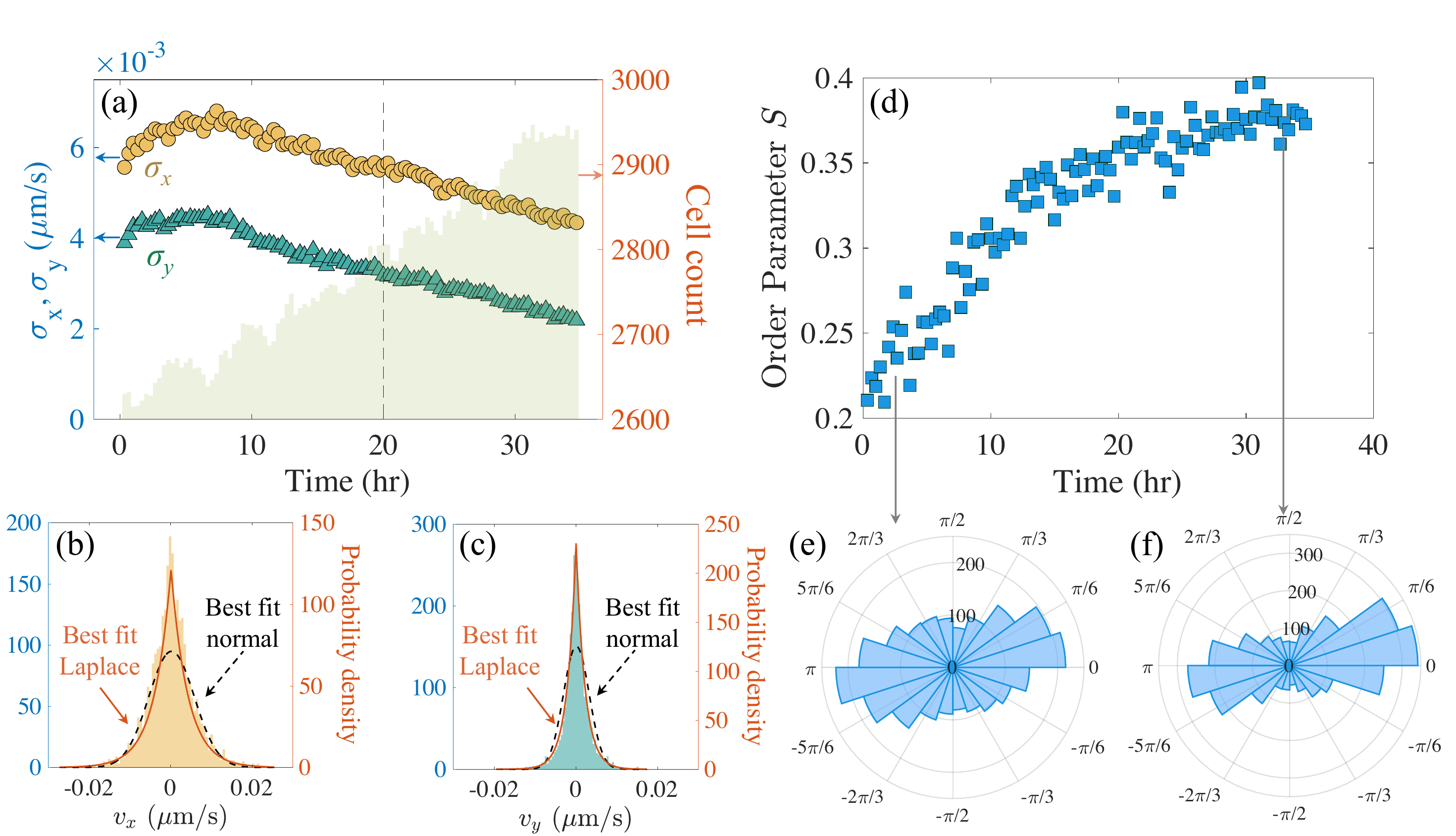}
   \caption{
   Parameters estimated from the data set. (a) The standard deviation of the cell velocities is shown in yellow circles and green triangles. The light-colored histogram denotes the cell number count (b-c). The velocity distribution at a representative time $= 20$ hours.  The dashed black lines denote the best fit normal distribution. The solid orange lines denote the best fit Laplace distribution. (d) The blue squares denote the polar order parameter, and the polar histograms of the velocity angle at two different time points are plotted in (e-f). }
\label{fig:velocity_alignment_eda}
\end{figure*}

From the experiment, we extract the 2D position vector  $\mathbf s_i(t)=(s_{i,x}(t), s_{i,y}(t))^T$ and velocity vector $\mathbf v_i(t)=(v_{i,x}(t), v_{i,y}(t))^T$ of cell $i$, at time frame $t$, for $i=1,...,n_t$, with $n_t$ being the number of cells on time frame $t$, for $t=1,...,T$. {The velocity is computed by dividing the displacement over time intervals as follows: 
$v_{i,x} = \frac{s_{i,x}(t+\Delta t)-s_{i,x}(t)}{\Delta t}$, and similarly for $v_{i,y}$. We then store the data matrix, where each row contains the 2D  position, velocity and a unique cell identification number resulting from trajectory linking. 
} The goal is to construct an interpretable  model and constrain the model by data, for reproducing the temporally dependent variability of the velocity and orientational order parameter. We plot the standard deviation of the velocity at $x$ and $y$ directions at each time frame  in Fig. \ref{fig:velocity_alignment_eda}(a). Since the nucleus does not distinguish between the head and tail, its order is apolar, 
we cannot apply the polar order parameter to quantify the system alignment \cite{lober2015collisions}. Instead, we first transform velocity angle  of the particle $i$ at time frame $t$, denoted by  $\theta_i(t)=\arctan(v_{i,y}(t)/v_{i,x}(t))$, to $[0,\pi]$ by letting: 
\begin{equation}
    \tilde \theta_i(t) = \begin{cases}
     \theta_i(t)+\pi & \text{if }  \theta_i(t)<0,\\
     \theta_i(t) & \text{if }   \theta_i(t)\geq 0.
    \end{cases}
    \label{equ:tilde_theta_i_t}
\end{equation}
The  orientational order parameter at time $t$ is an ensemble of transformed velocity angles amongst all cells: $S(t)= \langle \cos(2\tilde \theta_i(t))\rangle_i$.   
This way, antiparallelly traveling cell pairs ($\theta_i$ = $\theta_j + \pi$) have the same contribution to orientational order parameters. In Fig. \ref{fig:velocity_alignment_eda}(d), we show that the orientational order parameter $S(t)$ increases with time. 
The distributions of the velocity angles at two different time points are plotted in Fig. \ref{fig:velocity_alignment_eda}(e) and (f), showing that over time the velocity distribution becomes more parallel along $x$-direction.

To account for observed migrational patterns, we start by identifying unexpected deviations of the magnitude and distributions from the classical {Vicsek} Model. 
Exploratory data analysis (EDA)  \cite{tukey1977exploratory} is a useful step to visualize  patterns from complex  experimental data before performing  statistical tests and modeling. Here, we first perform EDA on various aspects of the data, followed by statistical tests and  estimation to identify features to include in the modeling. The main text focuses on the analysis of an experiment where cells initially cover roughly 50\% of the substrates. During the experiment, the order parameter increases rapidly with cell proliferation. We present results on the analysis of two additional experiments at different cell densities or on isotropic substrates in Appendix D.  

 \subsection{Permutation F-test on variances of the velocities}
 
To test whether the variances of the velocities along $x$ and $y$ directions are the same at all time frames, we first compute the sample standard deviation of the velocity vector at $x$ and $y$ coordinates: 
$\hat  \sigma^2_x(t)=\sum^{n_t}_{i=1} (v_{i,x}(t)- \bar v_x(t) )^2/(n_t-1) $  and  $\hat  \sigma^2_y(t)=\sum^{n_t}_{i=1} (v_{i,y}(t)- \bar v_y(t) )^2/(n_t-1) $,
where $\bar v_{x}(t)$ and $\bar v_{y}(t)$ are the mean of the velocity at time $t$ computed from an ensemble of all cells in the system, and both are close to zero. 
As shown in Fig. \ref{fig:velocity_alignment_eda}(a), we found that the standard deviation  of the velocity along the $x$ coordinate is always larger compared to that along $y$. The larger magnitude of cellular velocity along the $x$ coordinate is induced by the liquid crystal elastomer substrate, which is  molecularly aligned along the $x$ direction in this experiment. As the velocity magnitudes along $x$ and $y$ directions become more unequal over time, the orientational order parameter increases (Fig. \ref{fig:velocity_alignment_eda}(d)).

Is the observed difference between $\sigma_x(t)$ and $\sigma_y(t)$ due to signal or noise? Typically, testing the equality of the variance is performed using the F-test assuming both populations are normally distributed. 
However, as we will see later,  velocity distributions do not follow a normal (Gaussian) distribution, while the F-test is sensitive to the normality assumption \cite{box1953non}. 

Here we circumvent the normality assumption by running a permutation F-test. The permutation test is a general nonparametric test that is insensitive to the distributional assumption. 
At each time $t$, we begin by combining the velocities at $x$ and $y$ directions into a long vector. Subsequently, we randomly assign these combined velocity values into two groups of equal size, creating a permuted sample. We repeat this step $B$ times to collect  $B$ permuted samples:  $\mathbf v^{(b)}_x(t)=[ v^{(b)}_{1,x}(t),..., v^{(b)}_{n_t,x}(t)]$ and $\mathbf v^{(b)}_y(t)=[ v^{(b)}_{1,y}(t),..., v^{(b)}_{n_t,y}(t)]$  for $b=1,...,B$. Then we compute the permuted F-test statistics $F^{(b)}(t)={\hat  \sigma^{2}_{b,x}(t)}/ {\hat \sigma^{2}_{b,y}(t)}$, where  $ {\hat  \sigma^{2}_{b,x}(t)}$ and $\hat \sigma^{2}_{b,y}$ denote the sample variances of the velocity for $x$ and $y$ directions, respectively, from the permuted sample $b$.   The p-value is twice the minimum of the probabilities $\mbox{Pr}(F^{(b)}(t)>F(t))$ and $\mbox{Pr}(F^{(b)}(t)<F(t))$, which can be computed empirically using the permuted samples. 

\begin{figure}
    \centering
   \includegraphics[width=0.4\textwidth]{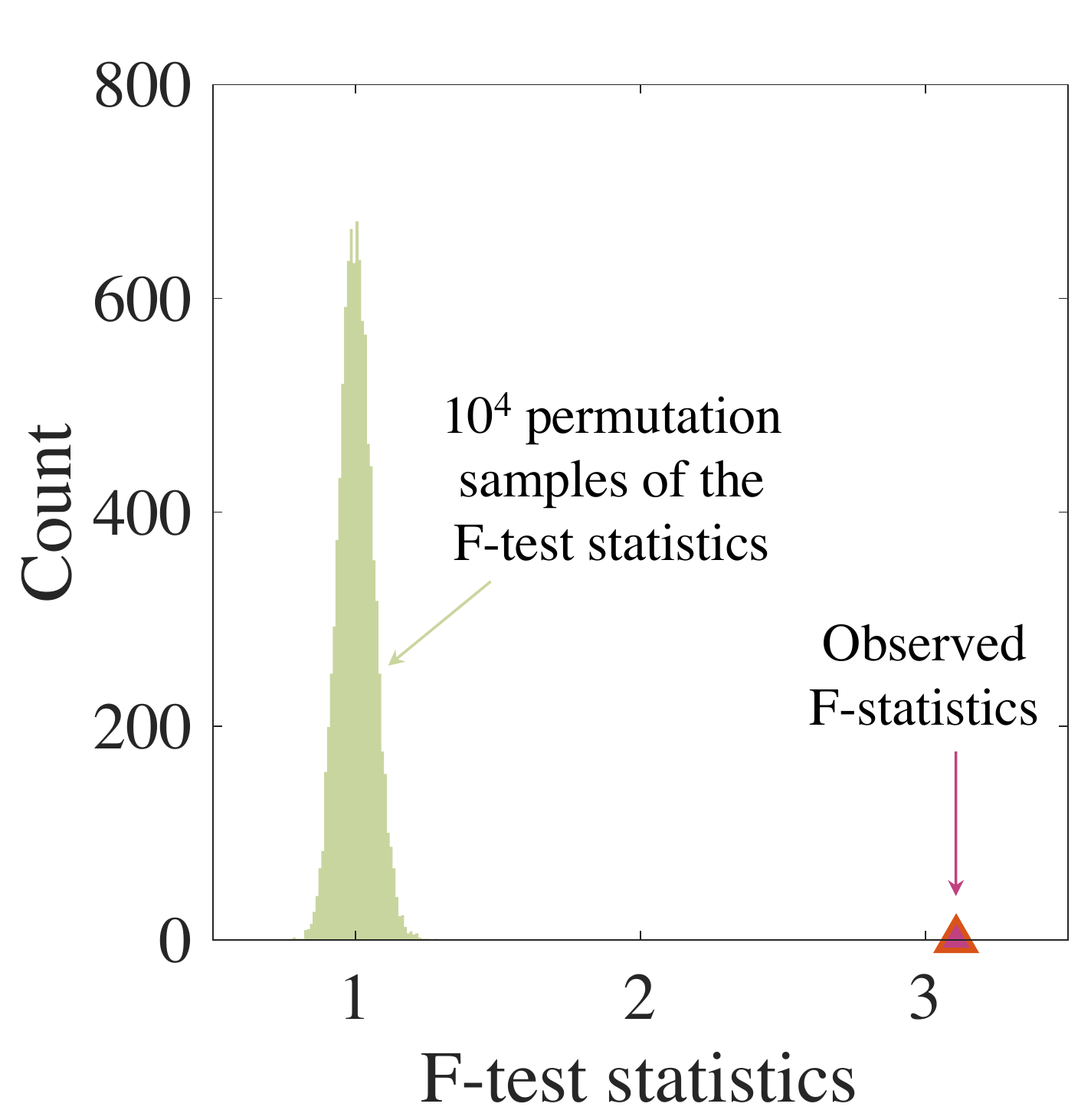}
    \caption{A permutation F-test shows that the variance of the fluctuation along $x$ and $y$ directions is  indeed unequal at time =20 hrs.}
    \label{fig:f_stat}
\end{figure}

Figure \ref{fig:f_stat}  shows the distribution of the F-test statistics of the permuted samples and the observed F-test statistics for velocity observations at time equal to 20 hours. {Since the observed F-test statistics is larger than any of the permutation sample}, the p-value  is smaller than $10^{-4}$, indicating that the variance of the velocity along  $x$ and $y$ directions is unequal at this time frame. We perform the permutation F-test for each of the time points and verify that the variances $\sigma^2_x(t)\neq \sigma^2_y(t)$ for all time points. Thus, cellular movements are anisotropic when cells are grown on a molecularly aligned substrate.

\subsection{Shapiro-Wilk test for normality of velocity distribution}
\label{subsec:non-Gaussian}

While at first glance cell motility bears much resemblance to passive, freely diffusing particles, our study reveals that velocity distributions in our system deviate from the Gaussian distribution typically observed for passive particles in a homogeneous environment. To illustrate that, we plot the probability density function of velocity distributions at a representative time (Fig. \ref{fig:velocity_alignment_eda}(b) and (c)).  Both velocity distributions along $x$ and $y$ have spiky modes and heavy tails, which cannot be captured by the best-fit Gaussian distribution (Fig. \ref{fig:velocity_alignment_eda}, black dashed curves), for which the mean and standard deviation are specified as the sample mean and sample standard deviation, respectively. The shape of the probability density distribution  suggests the use of the Laplace  distribution (Fig. \ref{fig:velocity_alignment_eda}(b) and (c), orange solid curves), which fits the velocity distributions reasonably well. 

\begin{figure}[t]
    \centering
   \includegraphics[width=0.45\textwidth]{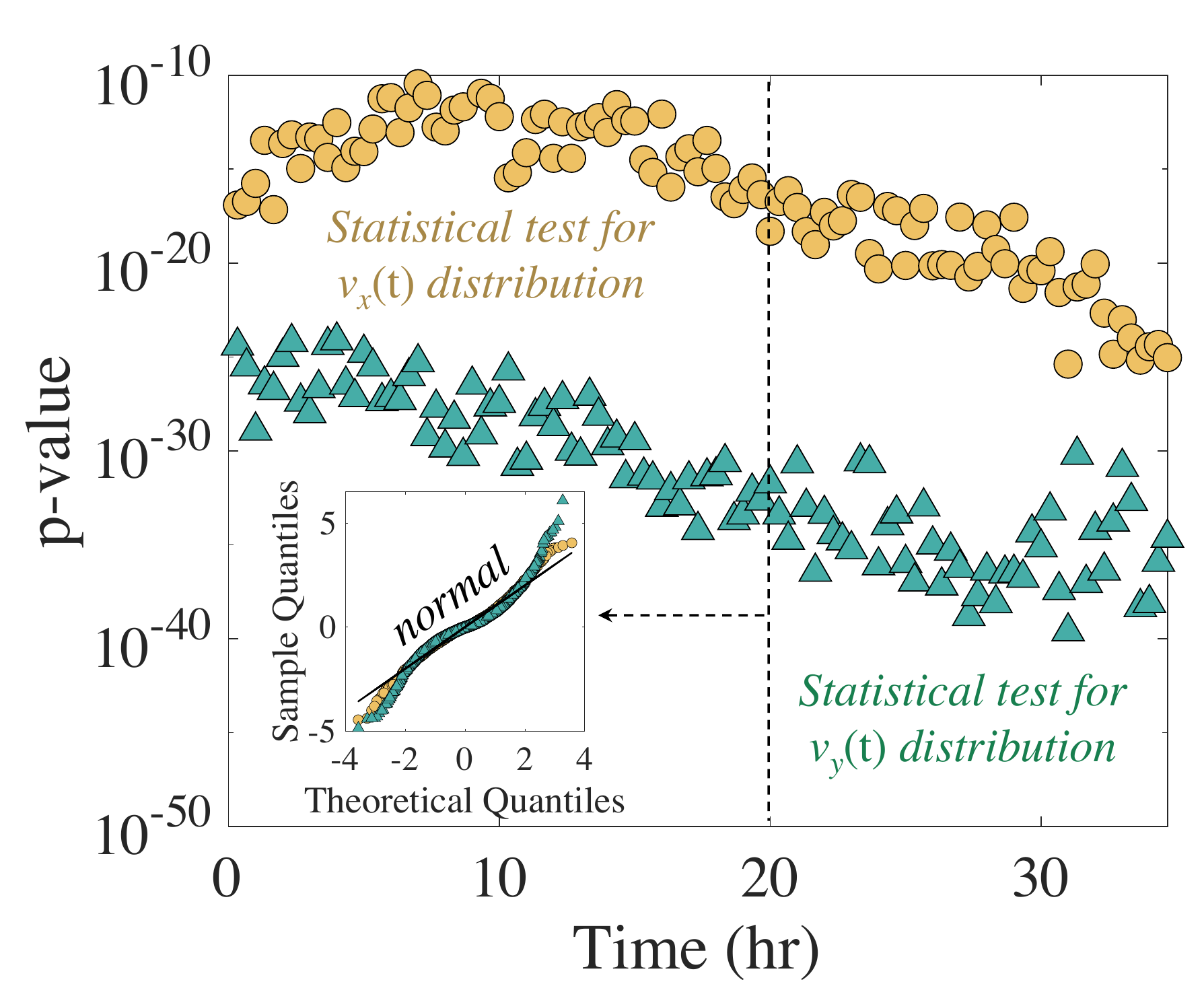}
    \caption{Shapiro-Wilk test for normality. The figure shows the p-value of individual frames, which is a statistical measurement applied to validate whether or not the observed data follows a normal distribution. A small p-value indicates statistical significance. The inset shows a representative quantile-quantile (QQ) plot at time = 20 hours to compare the sample quantiles of normalized velocity with the theoretical quantiles from the standard normal distribution. The black curve indicates the linear curve with slope 1 (normal). 
    }
    \label{fig:normality_test}
\end{figure}

\begin{figure*}[t] 
\centering
\includegraphics[width=\textwidth]{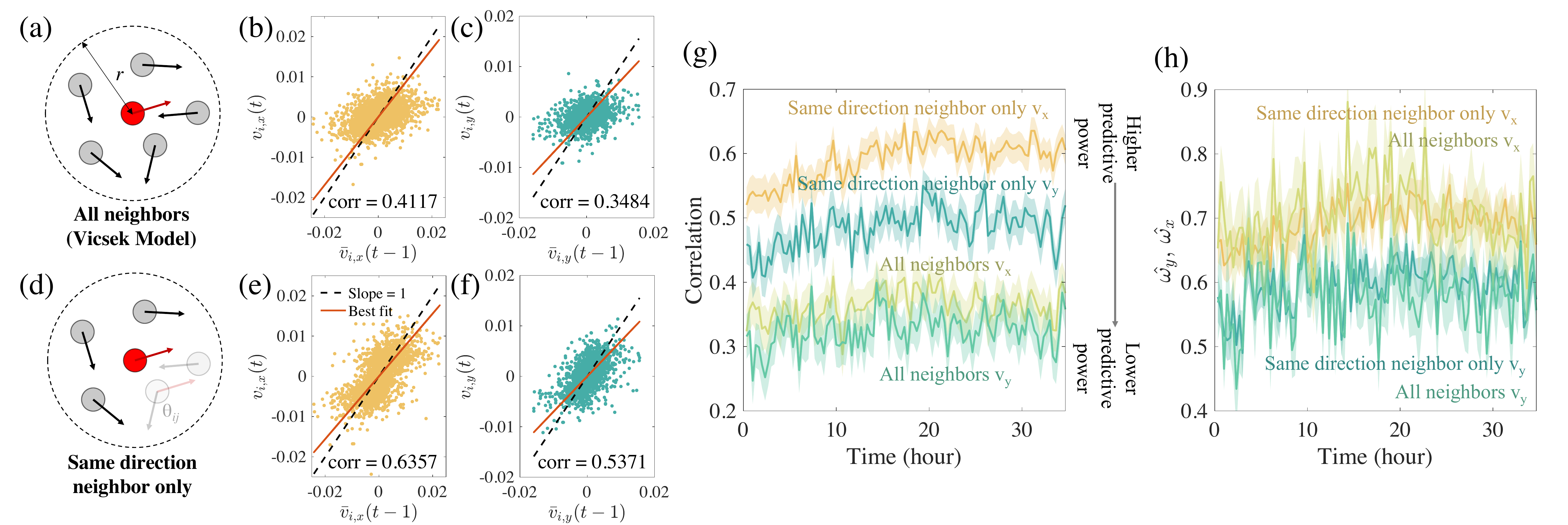}
   \caption{Effects of accounting for different neighbor ensembles. (a-c) Parameters and correlation plots using the neighbor ensemble of the classical Vicsek model for time = 20 hours.  (d-f)The scenario where only cells traveling in the same direction  are considered. [(b), (c), (e), and (f)] The correlation between the cell at $t$ versus the mean of the velocity of its neighbors at $t-1$. (g) Correlations between velocity $v_{i,x}(t)$, $v_{i,y}(t)$ of any cell $i$ and mean velocity of the neighboring cells from the previous step $\bar v_{i,x}(t-1)$, $\bar v_{i,y}(t-1)$, where the $95\%$ confidence  intervals, shown in fainted shades, are computed based on Fisher transformation of correlation coefficients. Two scenarios are shown, either by including all neighbors (bottom 2 curves), or only the same direction neighbors (top 2 curves). (h) The estimated weights, equivalent to the fitted  slopes, $\hat \omega_x(t)$, $\hat \omega_y(t)$ from these two scenarios and their confidence intervals are plotted.}  
\label{fig:same_direction_all_neighbors}
\end{figure*}

To test whether the velocity distribution follows the normal distribution at each time frame, we compute  p-values of the Shapiro-Wilk test for normality  \cite{shapiro1965analysis}, plotted in Fig. \ref{fig:normality_test}. The p-values are all extremely small, indicating the velocity distribution at any time frame substantially deviates from Gaussian. Furthermore, we plot the sample quantile-theoretical quantile from a normal distribution (QQ-plot) of the velocities along both directions at $20$ hours, both of which deviate from theoretical quantiles, shown as the black line in the inset.  The p-values get smaller at later time frames when the system approaches jamming, as the network effects are larger. Furthermore, the p-values for velocities along $y$-direction are  smaller than those from the $x$-direction, indicating that the deviation from normality is also bigger,  signaling more confinement along $y$-direction.

\subsection{Neighboring feature construction for cellular interaction}

Next, we discuss the contribution of  the neighbor ensemble to cell alignment. For any cell, the conventional choice of the neighboring set is to include all cells within a radial distance $r$ (as shown in Fig. \ref{fig:same_direction_all_neighbors}(a)).  Evidence of cell intercalation, where cells squeeze past their neighbors and they exchange positions, is increasingly found in recent studies \cite{guillot2013mechanics, bi2014energy}. When antiparallelly aligned cells move past each other, they minimally impact each other's velocity. This motivates us to investigate a neighboring set that excludes the cells with opposite velocities,  as shown in Fig. \ref{fig:same_direction_all_neighbors}(d). To compare these two approaches of accounting for the neighbors, in Fig. \ref{fig:same_direction_all_neighbors}(b), we plot the cellular velocity of each particle  $v_{i,x}(t)$ at time = 20 hours versus the mean velocity of neighboring cells at the previous time frame at $x$ coordinate, while in Fig. \ref{fig:same_direction_all_neighbors}(e) the same quantities are computed when the cells with opposite velocities are excluded from the neighboring set. As the neighboring set in the previous step includes the cell itself,  it is not surprising that both approaches yield a relatively high 1-step forecast accuracy, since individual cells tend to migrate with a persistent velocity \cite{rappel2017mechanisms}.

The correlation in Fig. \ref{fig:same_direction_all_neighbors}(e) is around $0.64$, which is substantially higher than the correlation of $0.41$ in  Fig. \ref{fig:same_direction_all_neighbors}(b), indicating that excluding the particles from the neighbors substantially improves the 1-step predictive accuracy. The same conclusion holds for velocity updates along the $y$ coordinate, shown in Fig. \ref{fig:same_direction_all_neighbors}(c) and (f). 
Furthermore,  
by evaluating the 1-step correlation over all time points (Fig. \ref{fig:same_direction_all_neighbors}(g)), we find that the method including only same-direction neighbors  substantially outperforms the one including all of the neighbors, as in the Vicsek model. 
These results indicate that the HDF cells in our experiment appear to distinguish between head and tail polarities \cite{li2018coordination}, despite extensive analogy that has been drawn between weakly interacting fibroblasts and active nematics \cite{duclos2014perfect,mueller2019emergence}, 

\subsection{Reduced magnitude of local alignment}

Unlike \cite{chate2008modeling},  we do not normalize the velocity to keep the velocity magnitude to be the same, as the  velocities can also change due to an increase in cell density.
Fig. \ref{fig:same_direction_all_neighbors}(e) and (f) 
indicate that  the velocity of the cell $i$ at time $t$ may be modeled by the mean of the velocity of the neighboring cells at time $t-1$, that is, $\mathbb E[v_{i,x}(t)]=w_x(t)  {\bar v}_{i,x}(t-1) $ and $\mathbb E[v_{i,y}(t)]=w_y(t)  {\bar v}_{i,y}(t-1) $, where $ {\bar v}_{i,x}(t-1) $ and $ {\bar v}_{i,y}(t-1) $ are the mean of the velocity of the neighboring cells at time $t-1$.  
To further test whether the statement is statistically significant, we compute the maximum likelihood estimator of the weights when the random fluctuation follows a Laplace distribution and generalized Gaussian distribution, with a different set of parameters at $x$ and $y$ coordinates. 

As shown in Fig. \ref{fig:same_direction_all_neighbors}(h), the maximum likelihood estimators of weights $w_x(t)$ and $w_y(t)$ are both smaller than 1. Furthermore, the shaded area shows that the 95\% confidence intervals of the estimation, which is calculated by the residual bootstrap estimate \cite{davison1997bootstrap}.  In all plots, we found the upper bounds of the $95\%$ confidence intervals of $w_x(t)$ and $w_y(t)$  are smaller than 1 at any time frame, indicating that the velocity of a cell aligns with the velocity of neighboring cells, while the velocity alignment is offset by other forces, such as frictional force between  substrates and cells  \cite{sepulveda2013collective,camley2016emergent,szabo2006phase,nagai2015collective}.

\subsection{Summary of data-driven findings and model development} 

{Here, we summarize the discovery from the EDA and statistical tests. First,  variances of the random motion along the $x$ and $y$ coordinates are distinct, and both decrease over time due to cell proliferation. Second, the probability density of the velocity distribution at both $x$ and $y$ directions has shown non-Gaussian behavior with a heavy tail and spiky mode near zero. Third, excluding cells with opposite velocities substantially improves the correlation between cellular velocity and mean velocity of the neighboring cells at the previous time step, as cells glide across each other in a manner reminiscent of intercalation \cite{huebner2018coming}. Fourth, the slope of the coefficient of a linear fit between the particle density and mean of neighboring particles in the prior time frame is smaller than one, indicating velocity alignment with its neighboring particles is offset by cell-substrate forces. All of the statistical tests we have developed so far are not restricted to the context of externally guided alignment; they are generally applicable to video microscopy which contains rich spatiotemporal data. Next, we illustrate how these analyses can be utilized to develop simulation models and parameter estimation for these models. 

}

 \section{Model constructed from selected features and maximum likelihood estimation}
 \label{sec:mle_model}
Our objective is to develop a minimum physical model that can account for  temporally dependent orientational  order parameters and velocities. 
We begin our model construction based on the seminal work by Vicsek \cite{vicsek1995novel}, consisting of agents moving at a constant speed and updating their direction of movement at each step. The ``new'' velocity direction is determined by the  summation of the average of the velocities of neighboring agents in the prior time frame, plus a random fluctuation. Despite its simplicity, this system exhibits a wide range of phenomena, including a transition from disordered to ordered behavior by adjusting the magnitude of the fluctuation.
Later modification to this model includes incorporation of distance-dependent interactions \cite{szabo2006phase}, 
which effectively reproduces observed phenomena such as local cohesion \cite{chate2008modeling} and jamming transitions \cite{henkes2011active}. These phenomena are characteristic of a migrating epithelium  \cite{mehes2014collective}, which has strong cell-cell junctions. 
Here we utilized a data-driven approach to make systematic improvements to the Vicsek model by incorporating the four selected features into the model. Though some of these aspects have been noted to some extent in literature, to our knowledge, there has been no work incorporating all of them and systematically testing their applicability for quantitatively reproducing the evolving orientational order observed in the experiment. 
The velocity vector of the $i$th particle at time frame $t$, $\mathbf v_{i}(t)=(v_{i,x}(t),v_{i,y}(t))^T$, for $t=1,...,T$ and $i=1,...,n_t$,  is modeled by two terms. The first and second terms represent the cell-cell, and cell-substrate interactions, respectively, as follows 
\begin{align}
v_{i,x}(t)&= w_x(t) \bar v_{i,x}(t-1)  + \epsilon_{i,x}(t), 
\label{equ:mv_1} \\
v_{i,y}(t)&=  w_y(t)  \bar v_{i,y}(t-1)+   \epsilon_{i,y}(t), 
\label{equ:mv_2}
\end{align}
where $\epsilon_{i,x}(t)$ and $\epsilon_{i,y}(t)$ are independent zero-mean random variables with variances $\mathbb V[\epsilon_{i,x}(t)]=\tau^2_x(t)$ and  $\mathbb V[\epsilon_{i,y}(t)]=\tau^2_y(t)$, respectively, $\bar v_{i,x}(t-1) $ and $\bar v_{i,y}(t-1) $ are the mean of the $x$ and $y$ directional velocities of neighboring particles at time $t-1$, and $w_x(t) $ and $w_y(t) $ are real-valued scalars denoting the weights. 
 The mean of the velocity at the $x$ and $y$ coordinates is modeled by 
 \begin{align}
 \bar v_{i,x}(t-1) = \frac{1}{p_{ne_i(t-1)}}{\sum_{j \in ne_i(t-1)} v_{j,x}(t-1)}, \\
  \bar v_{i,y}(t-1) = \frac{1}{p_{ne_i(t-1)}}{\sum_{j \in ne_i(t-1)} v_{j,y}(t-1)}, 
 \end{align} 
 where $ne_i(t-1)$ is the neighboring set of cell $i$ at time frame $t-1$ and 
 $p_{ne_i(t-1)}$ is the number of cells in the neighboring set. Let  $\mathbf s_i(t-1)$  be the 2D position of cell $i$ at time frame $t-1$. The neighboring set is defined as $ ne_i(t-1)=\{j: ||\mathbf s_{j}(t-1)-\mathbf s_{i}(t-1)
||<r \mbox{ and } \mathbf v_{j}(t-1) \cdot \mathbf v_{i}(t-1)>0 \}$, which contains  particles within a radius distance $r$ but excludes particles with opposite velocities. 
 
Section 2 \ref{subsec:non-Gaussian} provides compelling evidence that the fluctuation distribution in the model significantly deviates from Gaussian distributions. 
Here, we introduce two distributions to model the random fluctuations in velocity: the Laplace distribution (or the double exponential distribution) and the generalized Gaussian distribution (or the stretched exponential distribution).  Both distributions have been fit to the probability distribution of displacements that go beyond Gaussian assumptions, particularly when the system approaches jamming \cite{weeks2000three,chaudhuri2007universal,wang2012brownian,castillo2007local}. 
Yet, finding the maximum likelihood estimator  of the parameters with non-Gaussian fluctuations is not a computationally trivial task, given that the video contains $10^2$ time frames and each frame has over 2000 cells. Hence, we introduce  computationally-scalable approaches to compute the maximum likelihood estimator for models with these non-Gaussian fluctuations.

\subsection{Maximum likelihood estimator with Laplace fluctuation} 
For the sake of simplicity, we will only demonstrate the model using velocity along the $x$ direction as an example. The model for velocity along the $y$ direction can be constructed in a similar manner. 
To model the fluctuation  by the Laplace distribution  $\epsilon_{i,x}(t)\sim \mbox{Laplace}(0, \tau_x(t))$,  
entails that the residual of velocity of particle $i$ at time $t$   $e_{i,x}(t)=v_{i,x}(t)-w_x(t) \bar v_{i,x}(t-1) $ follows  
\begin{align}
    p(e_{i,x}(t)  \mid \tau_x(t))&=\frac{1}{\sqrt{2}\tau_x(t)}\exp\left(-\frac{\sqrt{2}|e_{i,x}(t)| }{\tau_x(t)}\right) . 
\end{align}

The maximum likelihood estimator of $w_x(t)$  can be obtained by iterative algorithms such as the expectation-maximization algorithm and iterative reweighted least squared algorithm \cite{schlossmacher1973iterative,phillips2002least}, while a faster alternative is through linear programming \cite{barrodale1973improved,barrodale1974solution}. The likelihood function and the maximum likelihood estimator  of the weight parameter,   $\hat w_x(t)$,  are introduced  in Appendix A. 
The maximum likelihood estimator of the standard deviation of the random fluctuation can be obtained by maximizing the profile likelihood after substituting in  $\hat w_x(t)$: 
\begin{align}
\hat \tau_x(t)&=\frac{\sqrt{2}}{n_t}\sum^{n_t}_{i=1} |v_{i,x}(t)-\hat w_x(t) \bar v_{i,x}(t-1) |. 
\label{equ:mle_tau_x_t}
\end{align}

\subsection{Maximum likelihood estimator with generalized Gaussian distribution}
In general, both the Laplace and the Gaussian distribution are special cases of the generalized Gaussian distribution (GGD), also referred to as a stretched exponential distribution \cite{mattsson2009soft,song2022microscopic}. Here, we illustrate the formulation by applying the model to fit the velocity component along the $x$ direction.  We assume that the residual $e_{i,x}(t)$ follows a zero-mean GGD with parameters $\alpha_x(t)$ and $\beta_x(t)$ at time frame $t$: 
\begin{align}
    &p(e_{i,x}(t)  \mid \alpha_x(t), \beta_x(t) )\nonumber \\
       =&\left( \frac{\beta_x(t)}{2\alpha_x(t)\Gamma\left(\frac{1}{\beta_x(t)}\right)}\right)\exp\left\{-\left(\frac{|e_{i,x}(t)|}{\alpha_x(t)}\right)^{\beta_x(t)} \right\}.  
   \label{eq:ggd}
\end{align}
The  variance of the GGD follows $ \tau^2_x(t)=\mathbb V[e_{i,x}(t)]=\alpha^2_x(t)\Gamma(3/\beta_x(t))/\Gamma(1/\beta_x(t))$, with $\Gamma(\cdot)$ denoting the gamma function.  When $\beta_x(t)=2$ and $\alpha_x(t)=\sqrt{2} \tau_x(t)$, GGD reduces to a  Gaussian distribution. When $\beta_x(t)=1$ and $\alpha_x(t)=\tau_x(t)/\sqrt{2}$, GGD reduces to a Laplace distribution. Therefore, the shape parameter $\beta_x(t)$ controls how close the GGD  resembles a Gaussian distribution. Similarly, the GGD  of the residual along $y$ can also be defined in terms of parameter  $\alpha_y(t)$ and $\beta_y(t)$ at any given time frame $t$. 

It should be noted that at each time frame, the  GGD has three parameters, and some of these parameters, such as the power parameter, are notoriously difficult to estimate \cite{green1984iteratively}. Hence, we derive closed-form derivatives to numerically compute the maximum likelihood estimator of the parameters, introduced in Appendix B. 

\section{Results}
Here we perform simulations to compare models with and without the selected features, and the simulation details are provided in Appendix C. 
\subsection{Model comparison by simulated orientational order parameter and velocity variability}

\begin{figure*}[t] 
\centering
\includegraphics[width=\textwidth]{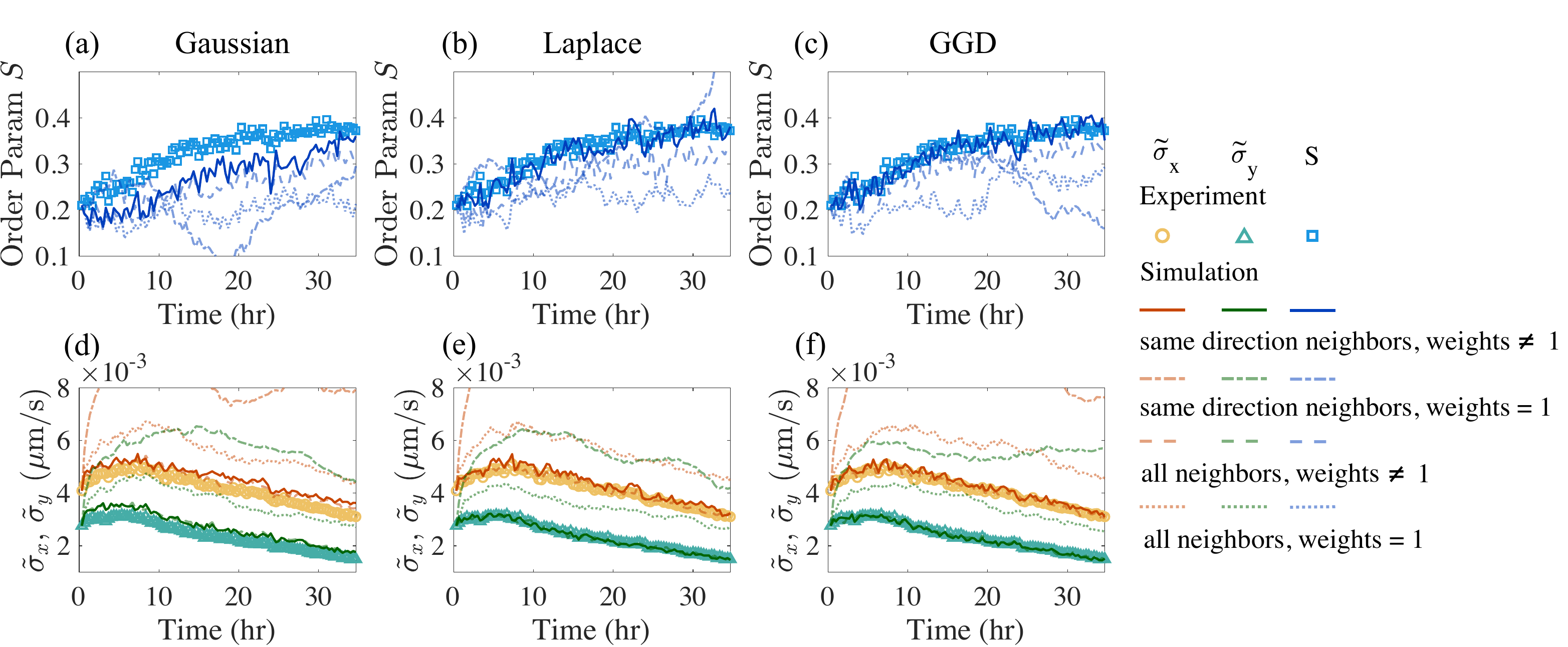}
   \caption{
 Reproducing orientational order parameters and mean absolute deviation of velocity with different simulation models at a representative radius $r$ = 75 $\mu$m. The left, middle, right panels compare simulations using the Gaussian, Laplace, GGD fluctuation with  experimental observations. For each type of fluctuation distributions, simulations of different interaction rules are presented.  
 (a-c) show the orientational order parameter, (d-f) show the average absolute deviation of the velocities. 
 }  
\label{fig:diff_sim_order_param_sigma}
\end{figure*}
We  compare models using simulated  order parameters and variability of velocity. The simulated  order parameters is computed by $S(t)= \langle \cos(2\tilde \theta_i(t))\rangle_i$ with $\tilde \theta_i(t)$ defined in Eq. (\ref{equ:tilde_theta_i_t}). As the velocity distribution more closely resembles a Laplace distribution, we use the average absolute deviation or $L_1$ loss to measure the variability of velocity: $\tilde \sigma_x(t)=\frac{1}{n_t}\sum^{n_t}_{i=1}|v_{i,x}(t) -\bar v_{x}(t)|$ and $\tilde \sigma_y(t)=\frac{1}{n_t}\sum^{n_t}_{i=1}|v_{i,y}(t) -\bar v_{y}(t)|$ with $\bar v_{x}(t)\approx 0$ and $\bar v_{y}(t)\approx 0$, instead of the more commonly used variance or $L_2$ loss to quantify the variability of the velocity. This is because the velocity distribution is closer to a Laplace distribution than a Gaussian distribution (Fig. \ref{fig:velocity_alignment_eda}(b) and (c)), and thus the $L_1$ loss is more appropriate to account for the variability of the distribution. 

In Fig. \ref{fig:diff_sim_order_param_sigma}, we compare  experimental observations and simulated results of the particle-based simulation, from which we computed the dynamics of the orientational order parameters and the average absolute deviation of the velocity. We apply the estimation and simulation procedure to scenarios where either all features are present or one or more of them are missing. Remarkably, we find simple models  from Eqs. (\ref{equ:mv_1})-(\ref{equ:mv_2}) with four features found in Section \ref{sec:feature_selection} can reproduce the dynamical progression of order parameters (solid curves in Fig. \ref{fig:diff_sim_order_param_sigma}(b) and (c)) and average absolute deviation of the velocity (Fig. \ref{fig:diff_sim_order_param_sigma}(e) and (f)), even though we do not fit a loss function for these ensemble properties directly. In Fig. \ref{fig:diff_sim_order_param_sigma_109} and Fig. \ref{fig:diff_sim_order_param_sigma_161} in Appendix D, we show the simulation using the estimated parameters from two other experiments and observe that our model is sufficiently general to capture the progression of system characteristics with varying initial densities and substrate materials.

In all simulation models, the magnitude and variability of the velocity along the $x$ coordinate are both larger than those along the $y$ coordinate, stemming from our first finding that the  velocity distribution is anisotropic. 
The growing anisotropy of the velocity induced by the molecularly aligned substrate leads to the increase of orientational order parameters. 

Second, the simulation with neighbor ensembles that include only cells traveling in the same direction (solid curves in Fig. \ref{fig:diff_sim_order_param_sigma}) more accurately reproduces both the progression of the orientational order and velocity than the simulation with interactions that includes all cells within a radial distance. 
This finding, which is derived solely from tracking the nuclei, reflects what has been observed from the video microscopy \cite{supplmat}: cells elongate, pulling on one another, as they migrate in the same direction. Conversely, when a pair of cells move pass each other in opposite direction, both cells maintain similar direction and velocities before the encounter. 
This effect likely occurs because of the complex interplay of cell-cell interactions, which are mediated through a cascade of signaling molecules. Consequently, the direction and polarity of contact are crucial factors in determining the strength of cell-cell interaction \cite{li2018coordination}. 

Third, the weight parameters $w_{x}(t)$ and $w_y(t)$ estimated by observations are always smaller than 1  (Fig. \ref{fig:same_direction_all_neighbors}(h)). In the simulation, we find that 
the model with  estimated weights   typically outperforms the model where the weights are constrained to be 1, particularly for reproducing the dynamical progression of the average absolute deviation of the velocity (Fig. \ref{fig:diff_sim_order_param_sigma}(a-c)). If the estimated slope parameters are equal to 1, the simulated velocity magnitude is frequently larger than what is observed.
In comparison, the inclusion of the estimated weights appears to resolve this issue, as the velocity is constrained. {This effect likely arises due to a loss of momentum to friction.} Furthermore, a unique aspect of our formulation is that the variability of velocity can change over time, applying estimated weights enables us to capture a wider array of behaviors, such as slow-downs. For instance, with increasing cell density, the average magnitude of the velocity must change. Ultimately, crowding leads to arrested dynamics \cite{angelini2011glass}. Nonetheless, simulation models often oversimplify this aspect by only modeling the velocity angle, overlooking the important role of slow-downs or heterogeneous dynamics \cite{vicsek1995novel,ginelli2010large}. 

Lastly, we find that the simulation model validates the finding that  velocities are not normally distributed (Fig. \ref{fig:normality_test})--the model with fluctuations following Laplace  distributions (Fig. \ref{fig:diff_sim_order_param_sigma}(b)) better reproduces the progression of the orientational order parameter than the model with Gaussian fluctuation (Fig. \ref{fig:diff_sim_order_param_sigma}(a)), even if they contain the same number of fitting parameters. Besides, the model with GGD fluctuation (Fig. \ref{fig:diff_sim_order_param_sigma}(c)) fits slightly better than the one with Laplace fluctuation, but it also contains one more parameter at each time point than both the Gaussian and Laplace distributions, and thus it has more flexibility in controlling the decay of the tail of the distribution. 
It is important to note that reproducing solely the variability of the velocity is inadequate to replicate the velocity orientational order parameter in our system, which is  sensitive to the change in the velocity distribution. 

\begin{figure}[t] 
\centering
\includegraphics[width=0.45\textwidth]{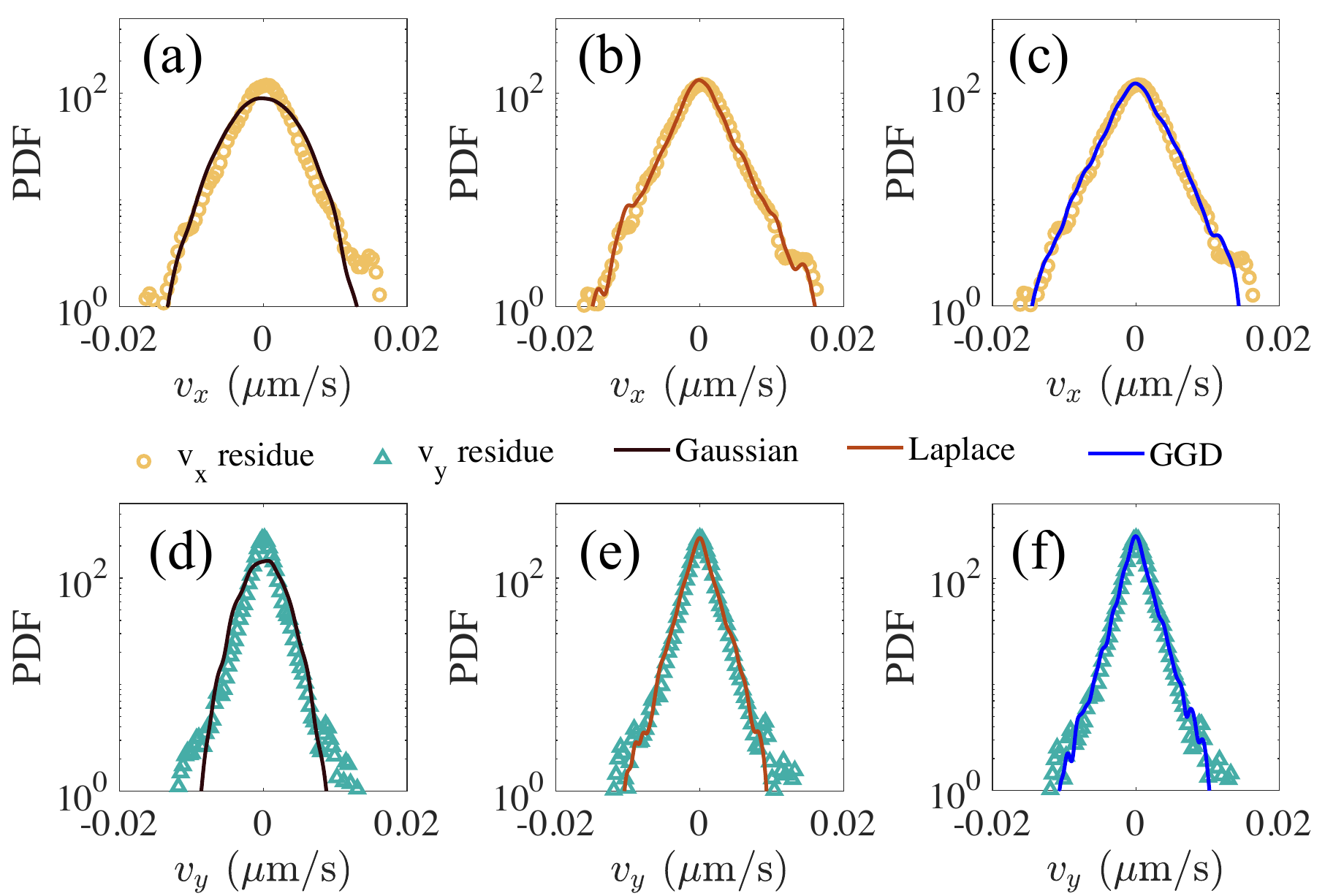}
   \caption{Distributions of random fluctuations between the observations and simulated models at a representative radius $r$ = 75 $\mu$m and time = 20 hours with the distributions of the fluctuation in the simulation following Gaussian (a,d), Laplace (b,e) and GGD (c,f).}
\label{fig:plot_density_Gaussian_Laplace_GGD}
\end{figure}

To further explore the difference between the effect of different random fluctuation distributions, 
in Fig. \ref{fig:plot_density_Gaussian_Laplace_GGD}, 
we show the distribution of the simulated residuals in $x$ direction in (a-c): $e^{sim}_{i,x} = v^{sim}_{i,x}(t)-\hat{w}_x(t) \bar{v}^{sim}_{i,x}(t-1)$ and in $y$ direction in (d-f): $e^{sim}_{i,y} = v^{sim}_{i,y}(t)-\hat{w}_y(t) \bar{v}^{sim}_{i,y}(t-1)$ by the colored circles and triangles, and the solid curve denotes the distribution of the residuals in the simulations. 
The fitted weight is derived from the maximum likelihood estimator where the fluctuations follow  Gaussian, Laplace, and GGD, from the left to the right. 
Simulated Gaussian fluctuation distributions along $x$ and $y$ directions are  shown in Fig. \ref{fig:plot_density_Gaussian_Laplace_GGD}(a) and (d), which underestimates the number of cells with near-zero velocities in both directions. Instead, simulation fluctuation distributions with either Laplace distribution or GGD better reproduce the experiments, as shown in Fig.  \ref{fig:diff_sim_order_param_sigma}(b), (c), (e) and (f).

\begin{table}
\caption{RMSE$_S$ for the orientational order parameter estimates. All assumed the weights parameters from the fluctuation from the $x$ and $y$ directions are separately estimated. The standard deviation of the velocity orientational order parameter is $0.050$ and an effective method (highlighted in bold)  has RMSE$_S$ smaller than this value. AN (= all neighbors) and SDN (= same direction neighbor) stand for methods with all neighboring cells within the radius and the same direction neighboring cells, respectively. Gau, Lap, and GGD are Gaussian, Laplace, and generalized Gaussian distributions of the fluctuation.}
\centering
\begin{tabular}{ c|c|c|c|c|c|c } 
 \toprule
Orientational order &  \multicolumn{3}{c|}{$\omega$ = 1}
  &  \multicolumn{3}{c}{$\omega$ $\neq$ 1}
  \\
   \cmidrule{2-7}
 parameter RMSE$_S$ & Gau & Lap &  GGD  & Gau& Lap &  GGD \\
 \midrule

{r = 25 $\mu$m AN}       & 0.17  & 0.12  & 0.14 & 0.090   & 0.061& 0.058 \\
{r = 25 $\mu$m SDN}      & 0.12  & 0.16   & 0.14 & 0.077   & \textbf{0.034} & \textbf{0.038} \\
\midrule
{r = 50 $\mu$m AN}       &  0.11 &  0.13 & 0.12 &   0.079 & \textbf{0.047 }& \textbf{0.043}\\  
{r = 50 $\mu$m SDN}      & \textbf{0.041}  & 0.12  & \textbf{0.033} &   0.069 & \textbf{0.022} & \textbf{0.025}\\
\midrule

{r = 75 $\mu$m AN}       & 0.13  &  0.11 &  0.10 &   0.081 & 0.058&\textbf{0.042}\\  
{r = 75 $\mu$m SDN}      & 0.15  & 0.060  & 0.099 &   0.063 & \textbf{0.020} & \textbf{0.018}\\

\bottomrule

\end{tabular}


\label{table:rmse_order_param}
\end{table}

\definecolor{Gray}{gray}{0.85}

\begin{table*}
\caption{RMSE for the average absolute  deviation of the velocity estimates $\tilde \sigma_x$ and $\tilde \sigma_y$. All assumed the weights parameters from the fluctuation from the $x$ and $y$ directions are separately estimated. The benchmark RMSE of the  average absolute  deviation (calculated by the standard deviation of the average absolute deviation)  for $\tilde \sigma_x$, $\tilde \sigma_y$ are 5.7$\times 10^{-4}$ and 5.3$\times 10^{-4}$, respectively.  An effective method (highlighted in bold)  has RMSE smaller than this value. AN (= all neighbors) and SDN (= same direction neighbor) stand for methods with all neighboring cells within the radius and the same direction neighboring cells, respectively. RMSE = (table value) $ \times 10^{-4}$.}
\centering
 \begin{tabular}{ l|c|c|c|c|c|c|c|c|c|c|c|c  } 
 \toprule
\multirow{4}{*}{\shortstack {Average absolute  deviation of \\ velocity estimates \\RMSE$_{\tilde \sigma_x}$, RMSE$_{\tilde  \sigma_y} {\times10^{4}}$ }} &  \multicolumn{6}{c|}{$\omega$ = 1}
  &  \multicolumn{6}{c}{$\omega$ $\neq$ 1}
  \\
   \cmidrule{2-13}
  
  &   \multicolumn{2}{c|}{Gaussian} & \multicolumn{2}{c|}{Laplace} &  \multicolumn{2}{c|}{GGD} &  \multicolumn{2}{c|}{Gaussian} & \multicolumn{2}{c|}{Laplace} &  \multicolumn{2}{c}{GGD}  \\
 {} & {$\tilde  \sigma_x$} & {$\tilde  \sigma_y$} & {$\tilde  \sigma_x$} & {$\tilde  \sigma_y$} & {$\tilde \sigma_x$} & {$\tilde  \sigma_y$} & {$\sigma_x$} & {$\tilde  \sigma_y$} & {$\tilde 
 \sigma_x$} & {$\tilde 
 \sigma_y$} & {$\tilde  \sigma_x$} & {$\tilde  \sigma_y$}\\
 \midrule

{r = 25 $\mu$m AN}      &  38 &  35 &  38 &  30 & 39 & 32 & \textbf{1.1}  & \textbf{2.7}  &  \textbf{1.4} &  \textbf{0.56} &   \textbf{1.3}  &    \textbf{0.67}  \\
{r = 25 $\mu$m SDN}     &  137 & 68 & 140  & 59  & 130 & 63 &  \textbf{3.0} &  \textbf{3.5} & \textbf{2.9}  &  \textbf{1.1} &   \textbf{2.6}  & \textbf{1.4} \\
\midrule
{r = 50 $\mu$m AN}      & 24 & 20  & 20  & 17  & 25  & 20 & \textbf{2.4}  &  \textbf{3.1}  &  \textbf{0.83} &  \textbf{0.52} &  \textbf{0.89}   & \textbf{0.54} \\  
{r = 50 $\mu$m SDN}     & 74 & 40  & 88  & 36  & 77 & 37 & \textbf{3.4}  &  \textbf{3.4} & \textbf{2.4}  &  \textbf{0.67} &   \textbf{1.7}  & \textbf{0.83} \\
\midrule

{r = 75 $\mu$m AN}      & 14  & 14  & 14  &  12 & 16  & 13 &  \textbf{3.0} &  \textbf{3.6} &  \textbf{0.87}  &  \textbf{0.47} & \textbf{0.72}    & \textbf{0.53} \\  
{r = 75 $\mu$m SDN}     & 42  &  34 & 52  & 32  & 49 & 32 &  \textbf{3.9} &  \textbf{3.4} & \textbf{2.1}  & \textbf{0.56}  &   \textbf{1.5}  & \textbf{0.51} \\

\bottomrule

\end{tabular}


\label{table:rmse_sigma}
\end{table*}

Here we use a neighbor radius of $r$ = 75 $\mu$m. {Fit of orientational order parameters by models with  different neighbor radii is plotted in Appendix E, and no significant difference amongst them has been observed. 
Thus, the radius seems to play a role in refining the fit, but to a much lesser degree than the choice of neighbor and fluctuation models.} 
To further explore difference between models, we present a summary of the results of incorporating various modeling parameters and compare the simulation to the experimentally obtained orientational order parameter in Table \ref{table:rmse_order_param}. The root mean squared error (RMSE) of the velocity orientational order parameter is calculated as the following:
\begin{equation}
    \mathrm{RMSE}_S = \sqrt{\sum_{t=1}^T\frac{( S_{t}^*- S_t)^2}{T}},
\end{equation}
where $S_{t}^*$ and $S_{t}$  are the  ensemble  orientational order parameters from the simulation and experiments at time frame $t$, respectively.    The RMSE for $\tilde \sigma_x$ and $\tilde \sigma_y$ can be defined similarly, and they are shown in Table \ref{table:rmse_sigma}. RMSE values that are better than the baseline average absolute  deviation are bolded. Values from Table \ref{table:rmse_order_param} and  \ref{table:rmse_sigma} indicate that, while there are several effective methods to describe the average absolute deviation of velocity, it is more difficult to capture the progression of the order parameter. This is not surprising as the order parameter is a complex function of the distribution, which is not explicitly controlled by any parameters in the likelihood function, given $v_{i,x}$ and $v_{i,y}$ are modeled independently. When modeling the velocity fluctuation, approaches incorporating Laplace and GGD fluctuations better capture the progression of the orientational order parameter, which is the main characteristic of the system, compared to the  model with the Gaussian fluctuation.

\subsection{Sensitivity analysis of the simulation}

 \begin{figure} 
\centering
\includegraphics[width=0.45\textwidth]{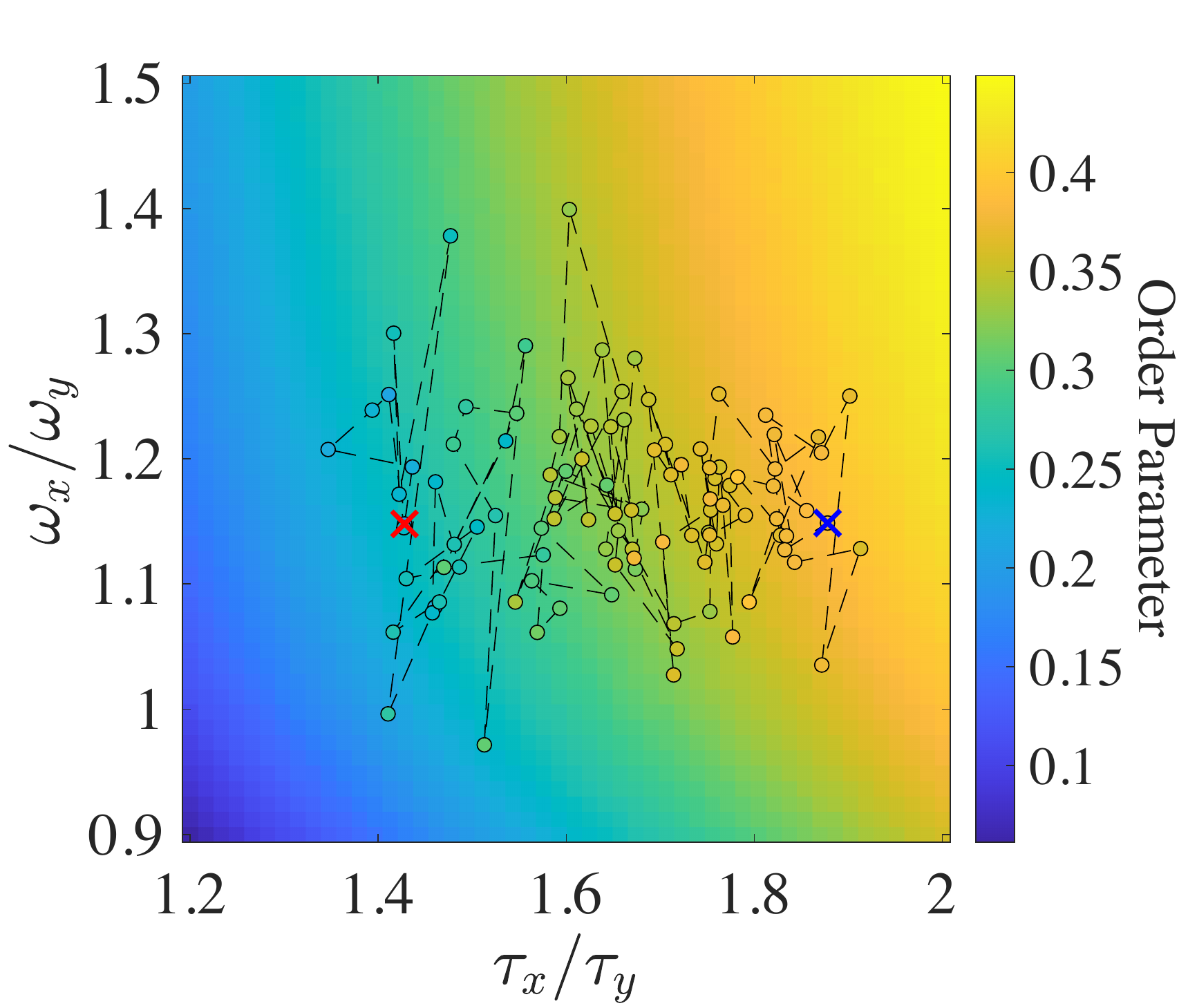}
   \caption{Change of order parameters due to the change of simulation parameters with $\tau_x = 0.02$ and $w_x=0.7$. The color bar corresponds to the simulated order parameter at long times. The dashed lines and circles denote the actual path of the experimental results plotted against the phase diagram. The colors of the markers correspond to the order parameter as denoted by the color bar. Red and blue X's denote the beginning and the end of the trajectory, respectively. } 
\label{fig:phase_diagram}
\end{figure}

As captured by Eqs. (\ref{equ:mv_1}) and (\ref{equ:mv_2}), the anisotropy of substrate materials induces asymmetric velocities in cells, which are manifested through both interactions and fluctuations. These contributions are temporally dependent, as changes in cell density due to proliferation lead to fluctuations in the velocity magnitude. For a given set of model parameters and fluctuation distribution, it will be helpful to understand their impacts on the asymptotic  order parameters, such as those that can be achieved after enough time has elapsed, for refining experimental designs and controls. 

Here, we perform a sensitivity analysis for simulation with Laplace fluctuations and all other selected features. We vary  the ratio of the variance of the fluctuation $\tau_x/\tau_y$ and the ratio of their weights $\omega_x/\omega_y$ in the model (Eqs. (\ref{equ:mv_1}) and (\ref{equ:mv_2})). While the variation of the order parameter is controlled by four parameters ($\tau_x$, $\tau_y$, $\omega_x$, and $\omega_y$), we find that the estimated $\omega_x$ (Fig. \ref{fig:same_direction_all_neighbors}(h)) does not change drastically over time. Furthermore, we show in Appendix B
that simulations with two different choices of $\tau_x$ do not have a large impact on the variation of the order parameter so long as the ratios of $\omega$'s and $\tau$'s remain the same. Hence, we are able to represent the asymptotic value of the order parameter on a 2D plot with $\tau_x=0.02$ and  $\omega_x=0.7$ fixed to be the mean of estimation over all time frames. 
Each simulation is implemented in 70 time points, where the order parameter fully equilibrates after 20 time points (see the evolution of the order parameter for select simulation parameter in Appendix B.
Due to this,  the order parameter in Fig. \ref{fig:phase_diagram} was computed by averaging the last 50 time points. 

We find that the asymptotic order parameter increases when either   $\tau_x/\tau_y$ or $\omega_x/\omega_y$ increases (Fig. \ref{fig:phase_diagram}). 
The order parameters computed from the experiment at different time points are overlaid on top of the diagram, where the warmer colors represent a larger magnitude of the order parameter, consistent with the color scheme of the phase diagram.
The ratio  $\omega_x/\omega_y$ fluctuates around a fixed value $\sim1.2$, potentially due to fixed properties of the external substrate guiding cells to migrate preferentially along $x$. That is to say, the substrate has a fixed modulus ratio along the $x$ and $y$ directions, as found in \cite{luo2022cell}. This ratio is reflected in the variance of the velocity along the corresponding directions, thereby restricting their ratio. As the system evolves over time, the caging effect becomes more prominent along $y$, which increases the ratio   $\tau_x/\tau_y$. 
This analysis provides further evidence to support our conclusion that substrate anisotropy and cell crowding both influence alignment. However, at low cell densities, the alignment effect is likely drowned out by noise.
Our findings also open up exciting design opportunities, where $\tau_x/\tau_y$ can potentially be controlled by varying the initial cell density or the  composition of the substrate \cite{martella2019liquid}, so the cell alignment dynamics can be tuned as a result. Overall, good agreements between the experiment and simulation are obtained.

\section{Discussions}

Cells can be induced to form aligned structures through a complex cascade of physiochemical signaling processes in both {\it in vitro} and {\it in vivo} settings. Experiments have shown that cells can migrate preferentially following the molecular orientation of the substrate. This preferential migration results from two interrelated effects:  cell-cell  and cell-substrate interactions. The former is due to the crowding due to the presence of other cells, and the latter is from cell polarization due to the substrate. Cell polarization occurs due to anisotropic interactions between the cell and the substrate or the extracellular matrix, including friction, damping, binding, and directional proliferation. 
Cell-cell and  cell-substrate interactions {both} drive the progression of velocity and order parameters. { Neither the  effect from one-body external potential acting on the random fluctuation term nor cell-cell interaction can be overlooked in our system. The combination of these two effects distinguishes our system and models from many prior studies \cite{stokes1991migration,wu2014three,selmeczi2005cell,bruckner2019stochastic}.}
We model these two contributions separately and quantitatively to reproduce the temporally dependent order parameters and velocity from the nonequilibrium process of cell alignment. 

{Here, we extract the cellular trajectory of a few thousand interacting cells from video microscopy,  and use this trajectory information to identify key factors required for reproducing the progression of velocities and  alignment order parameters. We found that when fibroblasts are cultured on an anisotropic substrate, global cellular alignment typically develops at high cell density but not low cell density \cite{luo2022cell}. This distinction sets our work apart from previous studies \cite{turiv2020topology,babakhanova2020cell}, as alignment forces acting on individual cells alone were  insufficient in inducing alignment. Thus,  our system cannot be simply regarded as particle alignment driven by a one-body potential from an external field, and the cell-cell interaction has to be included in modeling.   
To model complex cell-cell interactions and anisotropic random fluctuations,  we develop data-driven methods to select features and expand the baseline Vicsek model, by utilizing these selected features.} 
Our findings highlight the importance of anisotropic, non-Gaussian distributions of velocities  in reproducing the progression of cell movements guided by the molecularly aligned substrates.  Of equal importance is   the construction of neighboring interaction, by eliminating contributions from cells with opposite velocities and applying weights different from unity.  
These features are reminiscent of recent works that take into account the directionality of cell-cell instantaneous velocity in modeling interaction, particularly in the context of cell-extracellular matrix interaction \cite{zheng2020modeling}, and models that include contact-induced inhibition \cite{woods2014directional,zimmermann2016contact,camley2016emergent}.

The shape of the probability density of the temporally dependent velocity has a significant impact on accurately reproducing orientational order parameters and velocities through simulation. 
We observe that the experimental velocity distribution vastly deviates from a Gaussian distribution (Fig. \ref{fig:normality_test}),    
as the velocity distributions of a  large number of cells that have undergone minimal movement, along with a non-negligible group of cells that have moved a substantially long distance over a specific time interval $h$. 
Non-Gaussian distribution of displacement has previously been reported in systems with heterogeneous dynamics, as noted in \cite{leptos2009dynamics,slkezak2021diffusion}. 
The high concentration of immobile cells centered at zero can be attributed to cells that are confined by their neighbors. On the other hand, the presence of heavy tails of the velocity distribution can be attributed to the coordinated movements of many particles, also known as ``jumps'', as discussed in \cite{chaudhuri2007universal}. Empirical studies have also shown that a significant jump is often followed by other jumps \cite{wang2009anomalous}. Consequently, these probability densities exhibit slower rates of decay in the tail of the distribution compared to a Gaussian distribution, whereas they are more appropriately captured by the Laplace distribution \cite{wang2012brownian}. 
Although various experiments have shown anomalous distributions of displacement probability \cite{wang2009anomalous,wang2012brownian,slkezak2021diffusion,leptos2009dynamics,dieterich2008anomalous}, it has been less recognized that the distribution of random fluctuations plays a critical role in reproducing the progression of orientational order parameters observed in the experiments, partly because of a lack of means of efficient parameter estimation. Our work has filled in this gap by developing a maximum likelihood estimator for model parameters and a fast simulation scheme for non-Gaussian dynamics with the estimated parameters.  {Our findings demonstrate that the temporal progression of orientational order parameters can be more accurately captured when modeling cell-substrate interactions using a Laplace fluctuation of noise, as opposed to a Gaussian fluctuation, despite both having the same number of parameters. 
}

Our analysis has also revealed that the distinct behaviors of cells on isotropic or nematic substrates can be largely attributed to the presence or absence of anisotropic velocity. Our model captures this phenomenon by demonstrating that a difference in variability of the velocity along $x$ and $y$ directions leads to alignment (Figs. \ref{fig:diff_sim_order_param_sigma} and  \ref{fig:diff_sim_order_param_sigma_109}), while cells on isotropic substrates do not develop any order (Figs. \ref{fig:diff_sim_order_param_sigma_161}). We further show that asymptotic alignment can be controlled by the ratio of the weights and variance parameters (Fig. \ref{fig:phase_diagram}), and the order tends to approach zero as the velocity becomes more isotropic. 

Our procedure has several distinctive features that showcase data-driven discovery. 
A popular approach is to include a dictionary basis \cite{williams2015data,rudy2017data} and use regression to estimate the coefficients of this basis. However, including irrelevant features  is like adding unnecessary noise into the models, which can drastically reduce the efficiency of estimation. 
Rather than including all potential covariates, we begin by selecting the features to be included through exploratory data analysis and statistical tests. Such an approach produces more interpretable models and improves the efficiency of estimation, since only features with significant effects will be included. Second, {we provide a method for the feature selection and testing for cell studies. 
Conventionally comparing all models with a combination of $p$ features requires $2^p$ simulations, which could be very inefficient.   Here the statistical test of a feature  do not depend on assumptions of other features, and thus only $p$ tests are needed for feature selection.}
This new hybrid approach  can be utilized in other systems to aid physicists in  automating the tasks of visualization and feature selection, and extending baseline physics models by incorporating the selected features. 
 Lastly, by using the selected features we define a data generative model, instead of minimizing a loss function as conventionally adopted in other machine learning approaches. The data generative model provides a probabilistic mechanism, where the uncertainty of the estimation can be rigorously estimated and propagated throughout the analysis. 
 We demonstrate that the selected features are  key ingredients to capture the progression of alignment dynamics. 

A few additional directions are of interest for future work. 
First, it would be helpful to quantify  the effect and establish the physical mechanism of the selected features. These include, for instance, quantifying diminished influence of opposite-direction neighbors by morphological analysis of their deformations. 
It is also worthwhile exploring  fluctuation distributions with multiplicative noise variances (i.e. the noise variance of the fluctuation depending on individual cellular velocities), as these models were found to approximate the Laplace distribution in previous studies of cellular experiments \cite{selmeczi2005cell,wu2014three}. 
In addition, the role of imaging noise and tracking error will be  quantified. 

Second,  position-dependent interaction kernels were estimated using  observations of larger objects such as golden shiner \cite{katz2011inferring},  surf scoters \cite{lukeman2010inferring} and simulated particles \cite{lu2019nonparametric,gu2022scalable}.  It would be interesting to include a second interaction term with cellular positions as the inputs. Various extensions  of existing computational tools would be needed to achieve this goal, such as accelerating the estimation of kernel function with Laplace fluctuation distributions, and enabling temporally dependent interaction kernel functions. 
Furthermore, though the current model takes into account the influence of neighboring cells on velocity changes \cite{selmeczi2005cell,wu2014three}, it is important to note that cells also exhibit persistent motion as they travel, resulting in relatively smooth, one-dimensional spatial patterns \cite{rappel2017mechanisms}. 
{Characterizing this feature may require inferring second order statistics from observations, whereas a discretized model with first-order approximation may not be sufficient \cite{ferretti2020building}.}
On the other hand, the effect of increasing cell count and the corresponding slow-down as the system approaches jamming must be more carefully modeled in order to effectively forecast the variance of the velocity fluctuation.

Ultimately, the velocity changes are governed by forces \cite{brugues2014forces,alert2020physical,trepat2009physical}. Direct, in-situ force measurements will help further elucidate mechanisms and refine our models. Efforts must be made to reconcile the anisotropic nature of the substrate that induces cell alignment with the isotropic assumption often made in mechanical analysis to deduce force, as in the case of traction force microscopy \cite{sabass2008high}.
Other important considerations that can also be integrated 
are cell shapes and the restructuring of their subcellular structures such as actin filaments, which provide mechanical supports. When tracking cell migration, we largely rely on fitting an ellipse to the nuclei. In order to monitor the dynamics of aforementioned features, algorithms without particle tracking, such as Fourier-based differential dynamic microscopy  \cite{cerbino2008differential,giavazzi2009scattering,gu2021uncertainty}, can be applied to extract system properties such as mean squared displacements, for inspecting the mechanical properties of the system. {Finally,  in the case of other types of cells and microenvironments, similar procedures can be followed to discern the presence of different cell movement features in the model construction.}

\section{Acknowledgements}
The experimental portion of this work was supported by the Otis Williams Postdoctoral Fellowship from the Division of Mathematical, Life, \& Physical Sciences in the College of Letter \& Science, with partial support from the BioPACIFIC Materials Innovation Platform of the National Science Foundation under Grant No. DMR-1933487 (NSF BioPACIFIC MIP). MG acknowledges partial support from the National Science Foundation under Grant No. DMS-2053423. XF 
acknowledges the  support of the UC  Multicampus Research Programs and Initiatives (MRPI) program under Grant No. M23PL5990.  The authors thank Cristina Marchetti, Megan Valentine, and Matthew Helgeson for helpful discussions.

\bibliography{References_chronical_2022}
\section*{Appendices}
\setcounter{equation}{0}
\renewcommand{\theequation}
{A\arabic{equation}}

\section*{Appendix A maximum likelihood estimator with Laplace fluctuation}
\label{sec:laplace_mle}
Here, we discuss the maximum likelihood estimator  when the noise fluctuation follows the Laplace distribution. We use the observations for the $x$ coordinate as an example and the derivation for the $y$ coordinate follows similarly. We denote the observation vector $\mathbf v_x=(\mathbf v^T_{x}(1),\mathbf v^T_{x}(2),...,\mathbf v^T_{x}(T))^T$ with $\mathbf v_{x}(t)=(v_{1,x}(t),...,v_{n_t,x}(t))^T$ for time frame $t=1,...,T$, and assume that the initial velocity $ \mathbf v_x(0)$ is given. The likelihood function of the parameters $\mathbf w_x=(w_x(1),...,w_x(T))^T$ and $\bm \tau_x=(\tau_x(1),...,\tau_x(T))^T$ can be written as 
\begin{align*}
&p(\mathbf v_x\mid \mathbf w_x, \bm \tau_x, \mathbf v_x(0))  \\
=&\prod^T_{t=1}p(\mathbf v_x(t) \mid  \mathbf v_x(t-1),  w_x(t),  \tau_x(t) ) \\
=& \prod^T_{t=1} \prod^{n_t}_{i=1} p( v_{i,x}(t) \mid  \bar v_{i,x}(t-1),  w_x(t),  \tau_x(t) ) \\
=& (\sqrt{2}\tau_x(t))^{-\sum^T\limits_{t=1}{n_t} }\exp\left( -  \sum^T_{t=1}\sum^{n_t}_{i=1} \frac{\sqrt{2}\left|e_{i,x}(t)\right| }{{\tau_x(t)}} \right), 
\end{align*}
where the residual of the $i$th particle at time frame $t$ is defined as $e_{i,x}(t)= v_{i,x}(t)- w_x(t) \bar v_{i,x}(t-1) $.

For any time frame $t$, the maximum likelihood estimator of $w_x(t)$  is equivalent to  the least absolute deviation (LAD) regression below: 
\begin{align}
\hat w_x(t)=\underset{\tau_x(t),w_x(t)}{\mbox{argmin}}  \sum^{n_t}_{i=1}  |v_{i,x}(t)- w_x(t) \bar v_{i,x}(t-1)|  
\end{align}

Although there is no closed-form solution for the estimator in the LAD regression, fast algorithms, such as the  Barrodale and Roberts (BR) algorithm \cite{barrodale1974solution}, that can transform the LAD regression  into a linear programming problem, are available. The BR algorithm is available in standard software platforms. For instance, the package ``L1pack'' in the Comprehensive R Archive Network (CRAN) implements the BR algorithm to solve a LAD problem. After obtaining the estimator $\hat w_x(t)$, we substitute it into the likelihood function and maximize the profile likelihood to obtain the maximum likelihood estimator of $\tau_x(t)$, which is given in Eq. (\ref{equ:mle_tau_x_t}).

~~~

\section*{Appendix B maximum likelihood estimator with generalized Gaussian  fluctuation}

Next, we discuss the maximum likelihood estimator for models with generalized Gaussian  fluctuation. 
To do so, we denote three vectors of  parameters $\mathbf w_x$, $\bm \alpha_x$, and $\bm \beta_x$, each having $T$ dimensions. The logarithm of the likelihood function follows 
\begin{align*}
&\log(p(\mathbf v_x\mid \mathbf w_x,  \bm \beta_x,\bm \alpha_x))  \nonumber \\
= &\sum^T_{t=1} \left\{n_t\log\left( \frac{\beta_x(t)}{2\alpha_x(t)\Gamma\left(\frac{1}{\beta_x(t)}\right)}\right) -
 \sum^{n_t}_{i=1} \left(\frac{|e_{i,x}(t)|}{\alpha_x(t)}\right)^{\beta_x(t)}\right\},
\end{align*}
where  $e_{i,x}(t)= v_{i,x}(t)- w_x(t) \bar v_{i,x}(t-1) $ and $\Gamma(\cdot)$ denotes the Gamma function, with $w_x(t)\in \mathbb R$, $\alpha_x(t) \in \mathbb R^{+}$ and   $\beta_x(t) \in \mathbb R^{+}$.  

We first  differentiate the likelihood function with respect to $\alpha_x(t)$ and 
set it to zero. 
For any given $\beta_x(t)$ and $w_x(t)$, the likelihood is maximized when 
\begin{align}
    \hat \alpha_x(t)=\left(\frac{\beta_x(t)}{n_t} \sum^{n_t}_{i=1}|e_{i,x}(t)|^{\beta_x(t)} \right)^{\frac{1}{\beta_x(t)}}. 
    \label{equ:_hat_alpha_x_t}
\end{align}

After substituting the  $\hat \alpha_x(t)$ from Eq. (\ref{equ:_hat_alpha_x_t}) into the log-likelihood function, we obtain the logarithm of the profile likelihood of $(\bm w_x, \bm \beta_x)$:  
\begin{widetext}
\begin{align}
    \log(p(\mathbf v_x\mid \mathbf w_x, {\bm \beta_x}, \hat{\bm \alpha}_x)) 
    =\sum^T_{t=1} \left\{ n_t\log\left(\frac{\beta_x(t)}{2\Gamma(\frac{1}{\beta_x(t)})}\right)-\frac{n_t}{\beta_x(t)}\log\left(\frac{\beta_x(t)}{n_t} \sum^{n_t}_{i=1} |e_{i,x}(t)|^{\beta_x(t)}\right)-\frac{n_t}{\beta_x(t)} \right\}.
    \label{equ:profile_lik_w_alpha}
\end{align}
\end{widetext}
Denoting the logarithm of the profile likelihood by  $\ell( \mathbf w_x, \bm \beta_x)=\log(p(\mathbf v_x\mid \mathbf w_x,{\bm \beta_x},  \hat{\bm \alpha}_x))$ and differentiating Eq. (\ref{equ:profile_lik_w_alpha}) with respect to $w_x(t)$ and $\alpha_x(t)$, we then find
\begin{widetext}

\begin{align}
\frac{\partial\ell( \mathbf w_x, \bm \beta_x)}{\partial w_x(t)}&=\frac{n_t \sum^{n_t}_{i=1}|e_{i,x}(t)|^{\beta_x(t)-1} \bar v_{i,x}(t-1)\mbox{sgn}(e_{i,x}(t))}{\sum^{n_t}_{i=1}|e_{i,x}(t)|^{\beta_x(t)}}, \label{equ:dev_w_x_t} \\
\frac{\partial\ell( \mathbf w_x, \bm \beta_x)}{\partial \beta_x(t)}&=\frac{n_t}{\beta_x(t)}+\frac{n_t}{\beta^2_x(t)} \Psi\left(\frac{1}{\beta_x(t)}\right)+\frac{n_t}{\beta^2_x(t)}\log\left(\frac{\beta_x(t)}{n_t}\sum_{i=1}^{n_t}|e_{i,x}(t)|^{\beta_x(t)}\right)-\frac{n_t}{\beta_x(t)}\frac{\sum_{i=1}^{n_t}\left(|e_{i,x}(t)|^{\beta_x(t)}\log|e_{i,x}(t)|\right)}{\sum_{i=1}^{n_t}|e_{i,x}(t)|^{\beta_x(t)}}, \label{equ:dev_beta_x_t} 
\end{align}
\end{widetext}
where $\Psi(z)=\Gamma'(z)/\Gamma(z)$ stands for the ratio  between the derivative of a Gamma function and a Gamma function for any $z$, and $\mbox{sgn}(e_{i,x}(t))$ denotes the sign of $e_{i,x}(t)=v_{i,x}(t)- w_x(t) \bar v_{i,x}(t-1) $. Thereafter, we  iteratively maximize the likelihood function  with respect to $w_x(t)$ and $\beta_x(t)$ using the profile likelihood in Eq. (\ref{equ:profile_lik_w_alpha}) and closed-form derivative in Eqs. (\ref{equ:dev_w_x_t}) and (\ref{equ:dev_beta_x_t}) by the  low-storage quasi-Newton
optimization method (L-BFGS) \cite{nocedal1980updating}. Note that, for each $t$, we only need to iteratively maximize the log profile likelihood with respect to two parameters, making the computational procedure both fast and robust.

\begin{figure}[h] 
\centering
\includegraphics[width=0.48\textwidth]{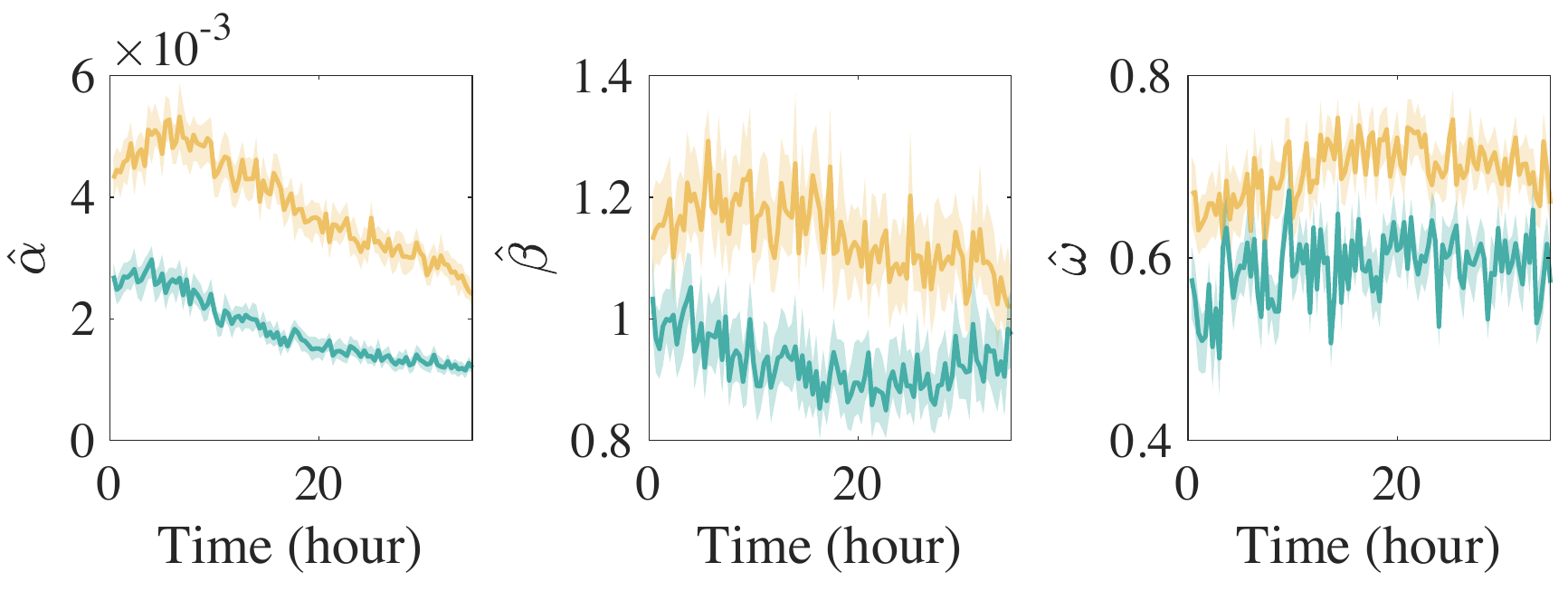}
   \caption{Parameter estimation with interval (shaded) for a Generalized Gaussian distribution following Eq. (\ref{eq:ggd}) along $x$-direction (yellow), and along $y$-direction (green).} 
\label{fig:param_interval}
\end{figure}
The maximum likelihood estimators of parameters $\alpha_x(t)$,  $\alpha_y(t)$, $\beta_x(t)$,  $\beta_y(t)$, $w_x(t)$ and  $w_y(t)$ assuming the fluctuation follows the generalized Gaussian distribution, are shown  in Fig. \ref{fig:param_interval}. The estimated $\beta_x(t)$ and $\beta_y(t)$ are both close to 1, which means that the fluctuation is closer to the Laplace distribution. In addition, $\beta_y(t)$ is typically smaller than  $\beta_x(t)$, as there is more confinement in the $y$ direction because of the substrate-imposed directionality. Furthermore, both $w_x(t)$ and $w_y(t)$ are smaller than 1 for any $t$,  consistent with the findings when  assuming the fluctuation follows the Laplace distribution.

\begin{figure}[t] 
\centering
\includegraphics[width=0.45\textwidth]{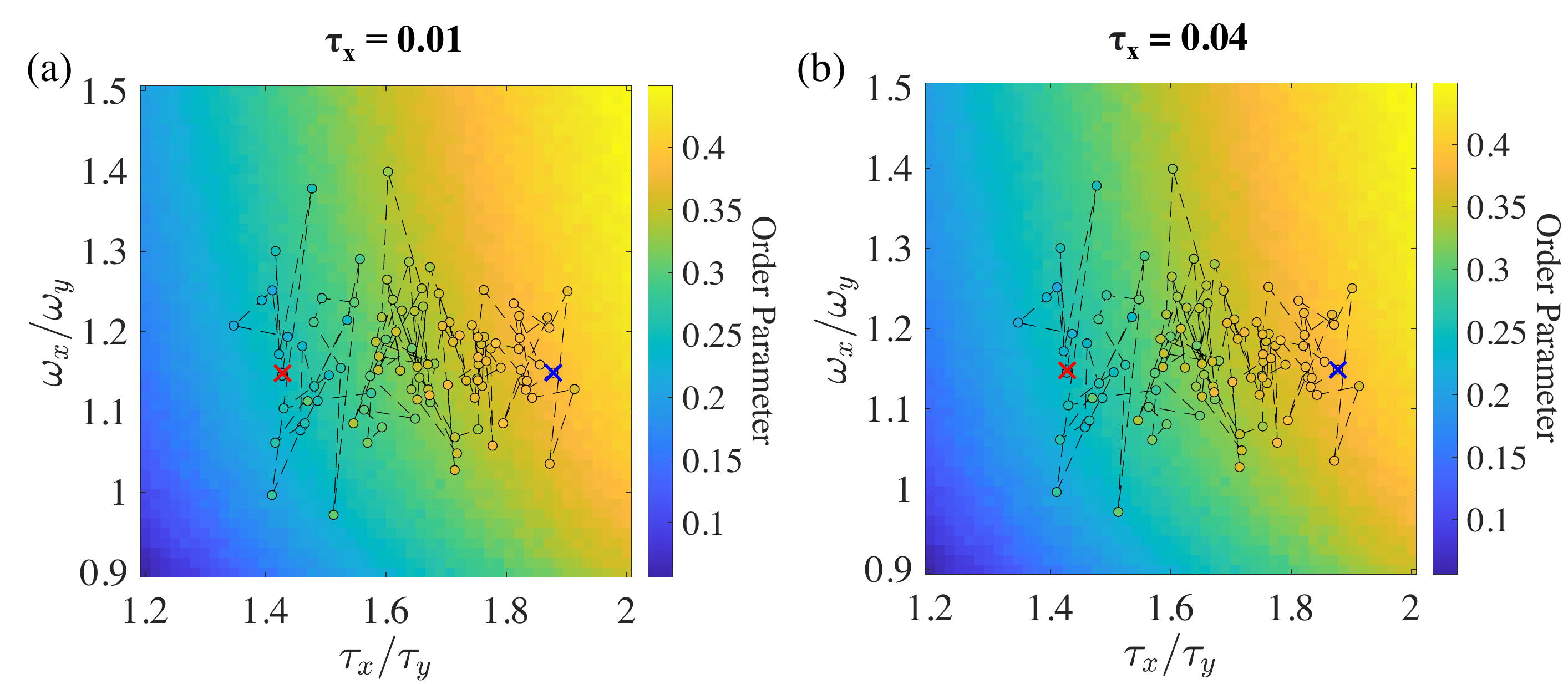}
   \caption{(a-b) Change of order parameters due to the change of simulation parameters with $\tau_x$ = 0.01 in (a) and $\tau_x$ = 0.04 in (b), which cover the range of the observed $\tau$'s. $w_x=0.7$ are assumed for both cases. Red and blue X's denote the beginning and the end of the trajectory.} 
\label{fig:phase_diagram_appendix}
\end{figure}

\begin{figure}[t] 
\centering
\includegraphics[width=0.42\textwidth]{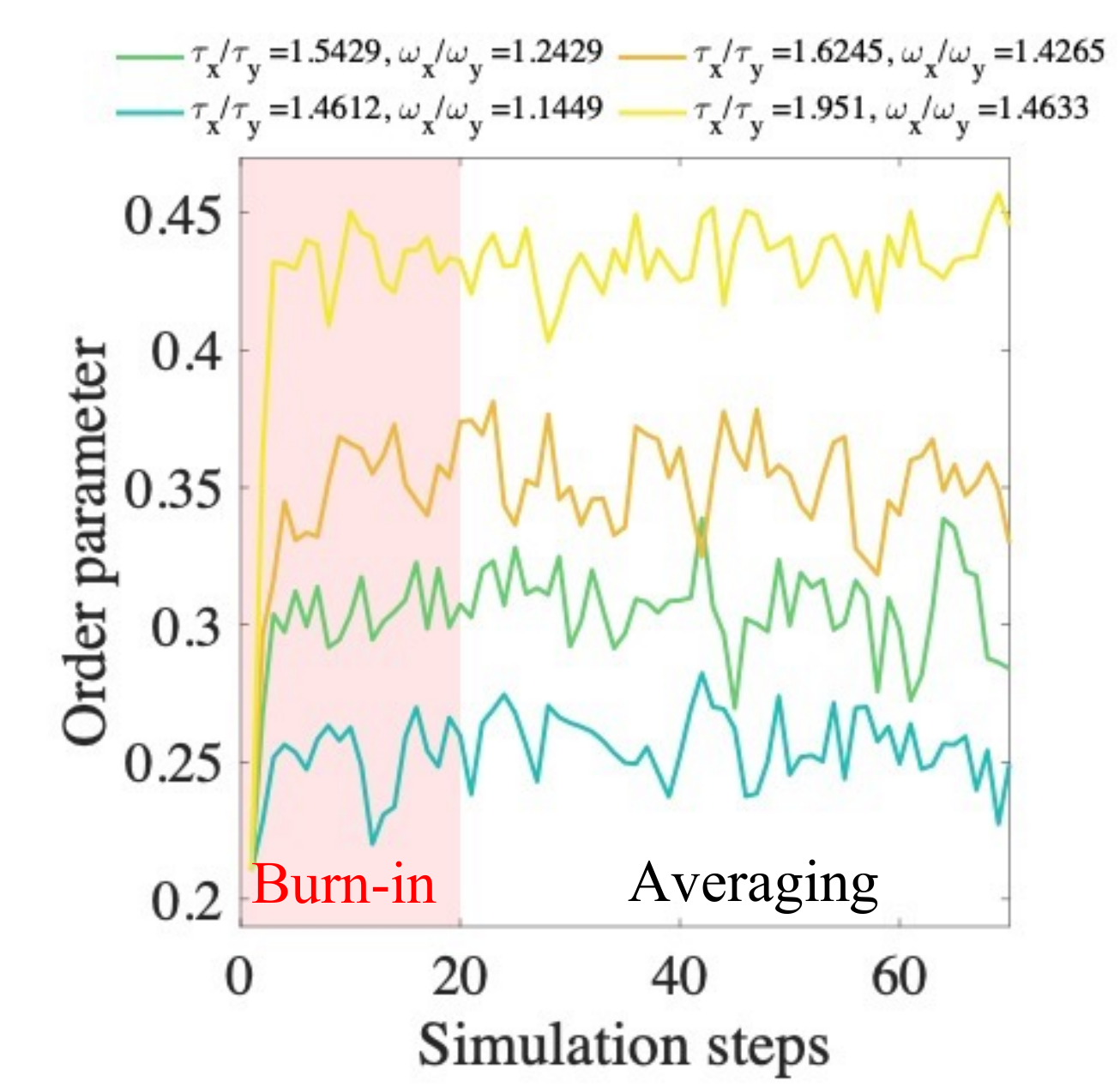}
   \caption{Several typical simulation progressions, where the first 20 steps are regarded as  burn-ins and left out in computing the averages in the phase diagram. } 
\label{fig:burn_in}
\end{figure}

To demonstrate the impact of parameter selection on the order parameter, we investigate the 2D parameter space defined by the ratios $\omega_x/\omega_y$ and $\tau_x /\tau_y$. Figure \ref{fig:phase_diagram} illustrates one possible path of order development, using a selected set of parameters $\tau_x$=0.01 and $\omega_x$=0.7. Next, we show that when $\tau_x$=0.01 and $\tau_x$=0.04, encompassing the experimentally observed values range, the order parameter exhibits consistent behavior along the temporal course of the experiment for different $\tau_x$ values (= 0.01, 0.02, 0.04) [Fig. \ref{fig:phase_diagram_appendix}(a) and Fig. \ref{fig:phase_diagram_appendix}(b)]. To determine the order parameter, we average the last 50 steps in the simulation, as it reaches a plateau after the first 20 steps (Fig. \ref{fig:burn_in}).

\section*{Appendix C Details on simulation}
\label{sec:simulation_details}
We first briefly introduce the setup of the simulation:
 Cells are modeled as particles, and their initial velocities and positions are imported from the data. For simplicity, we assume that the cell number remains constant.
 We start by partitioning the space into grids, and the spacing between gridlines is no smaller than the cell-cell interaction radius $r$.
 In our simulation, grids are populated with cells based on cells' positions $\mathbf s_i(t)=(s_{i,x}(t), s_{i,y}(t))^T$ (Fig. \ref{fig:overview}(e)). The grid-based approach { \cite{ginelli2016physics}} improves computational efficiency {for searching neighbors} in simulation  since as such, identifying a cell's neighbors only requires searching the nine grids surrounding it. Using this method,  computing the likelihood function for parameter estimation and simulation only requires $\mathcal{O}(\sum_{t=1}^{T} n_t)$ operations, which is much faster than the conventional approach requiring $\mathcal{O}(\sum_{t=1}^{T} n^2_t)$ operations, where $n_t$ is the number of cells at time frame $t$. {We note that this coarse-graining strategy to search for neighbors only improves computational efficiency, but does not change the simulation results.}

\begin{figure*}[t] 
\centering
\includegraphics[width=.98\textwidth]{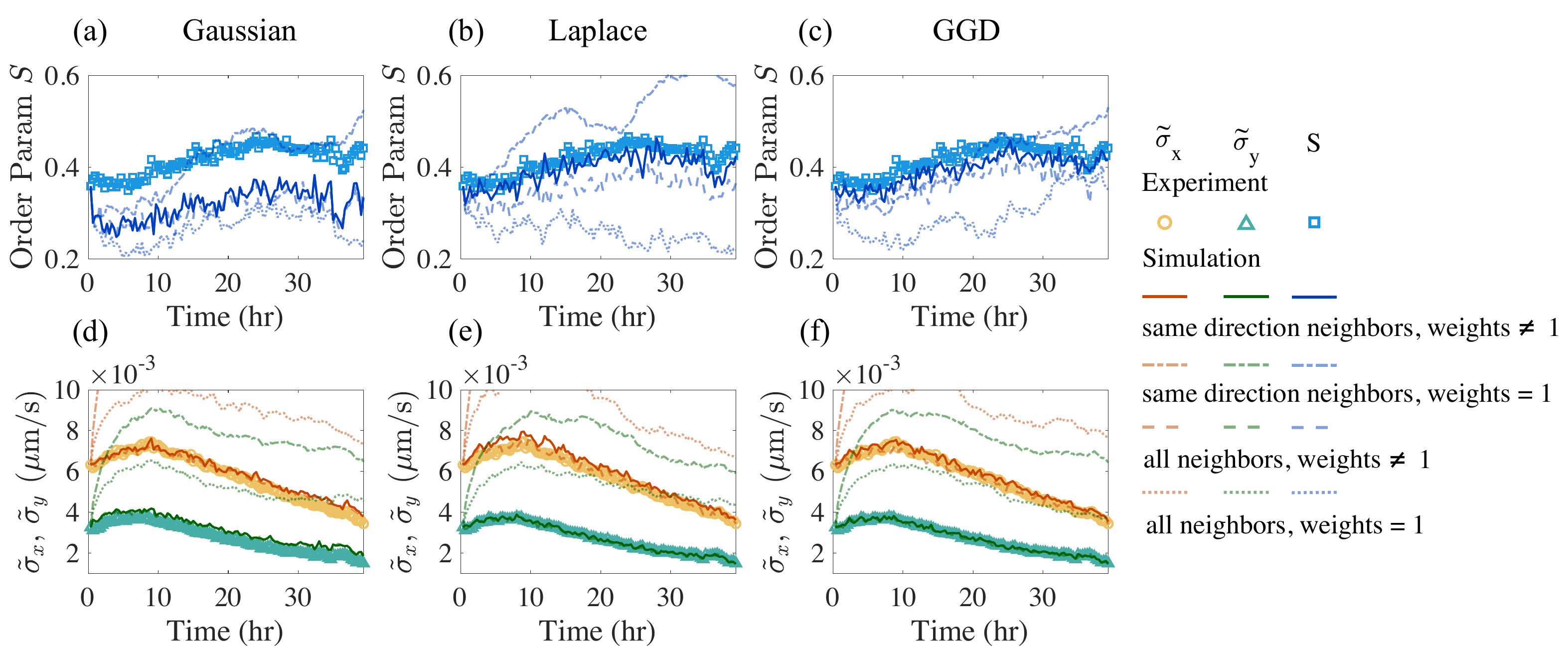}
   \caption{
   Reproducing orientational order parameter (a-c) and average absolute deviation of velocities (d-f) from different simulation models at a representative radius $r$ = 75 $\mu$m for experiments of cell migrating on a nematic substrate when starting at a higher initial density.} 
\label{fig:diff_sim_order_param_sigma_109}
\end{figure*}

 The velocities $v_{i,x}(t)$ and $v_{i,y}(t)$ are updated according to Eqs. (\ref{equ:mv_1})-(\ref{equ:mv_2}), with three types of distribution of fluctuations due to the cell-substrate interaction: Gaussian, Laplace, or GGD; two types of neighbors: interactions with all neighboring cells or only neighboring cells with the same direction velocity, and weights ($\omega_x$ and $\omega_y$) of the interaction fixed to be 1 or estimated from the data, which are due to cell-cell interactions. All parameters are estimated based on the maximum likelihood estimators introduced in Section \ref{sec:mle_model}.

\section*{Appendix D Additional results on experimental data}

\begin{figure*}[t] 
\centering
\includegraphics[width=\textwidth]{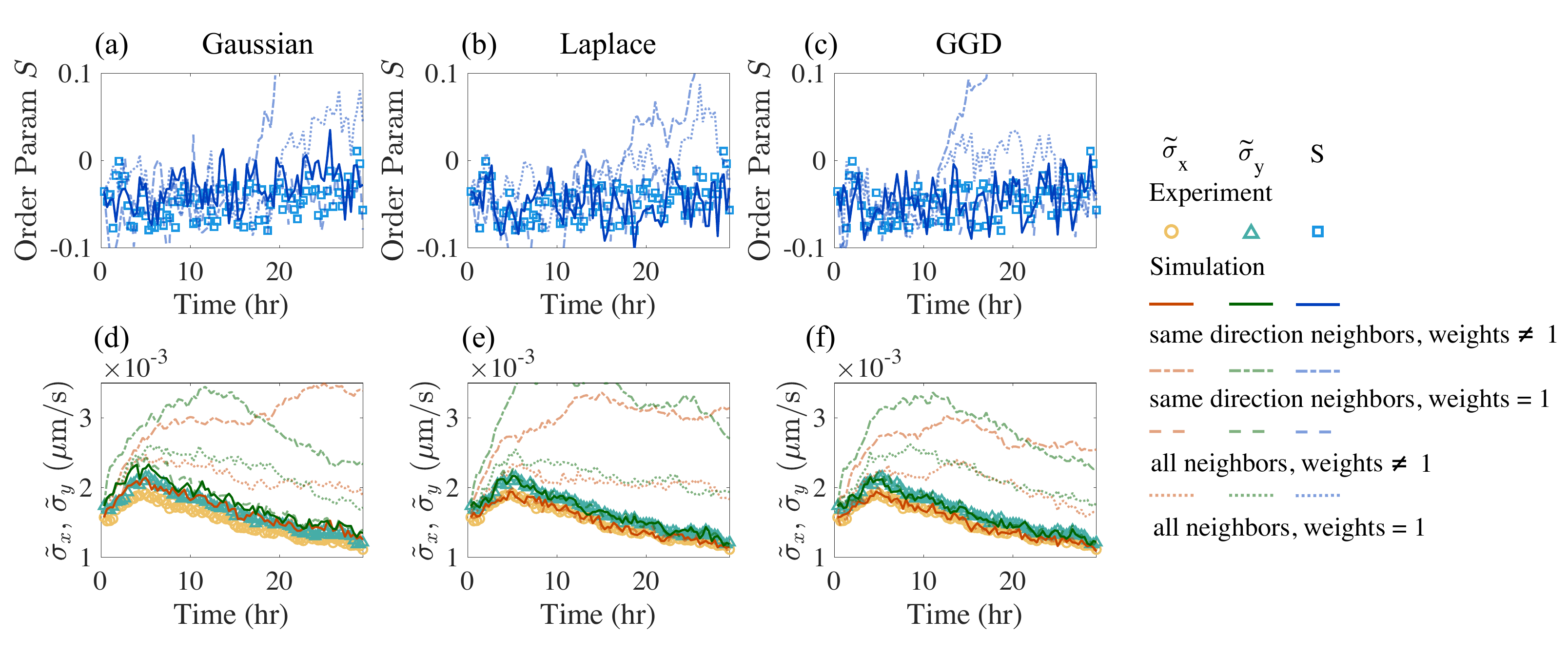}
   \caption{ 
   Reproducing orientational order parameter (a-c) and average absolute deviation of velocities (d-f) with different simulation models at a representative radius $r$ = 75 $\mu$m for experiments of cells moving on an isotropic substrate, which do not align, and have isotropic velocity.} 
\label{fig:diff_sim_order_param_sigma_161}
\end{figure*}

 The four new features are used to construct a minimum physics model that can quantitatively capture the progression of the orientational order parameter and velocity distribution, for various experiments with different initial cell densities and liquid crystal elastomer substrates. The results of several additional experiments are also analyzed to validate the generality of our algorithm (Fig. \ref{fig:diff_sim_order_param_sigma_109}-\ref{fig:diff_sim_order_param_sigma_161}). Similar to the main text, we evaluate  models by velocity orientational order parameter and the average absolute deviation of the velocity at each time point.

\begin{figure*}[t] 
\centering
\includegraphics[width=0.8\textwidth]{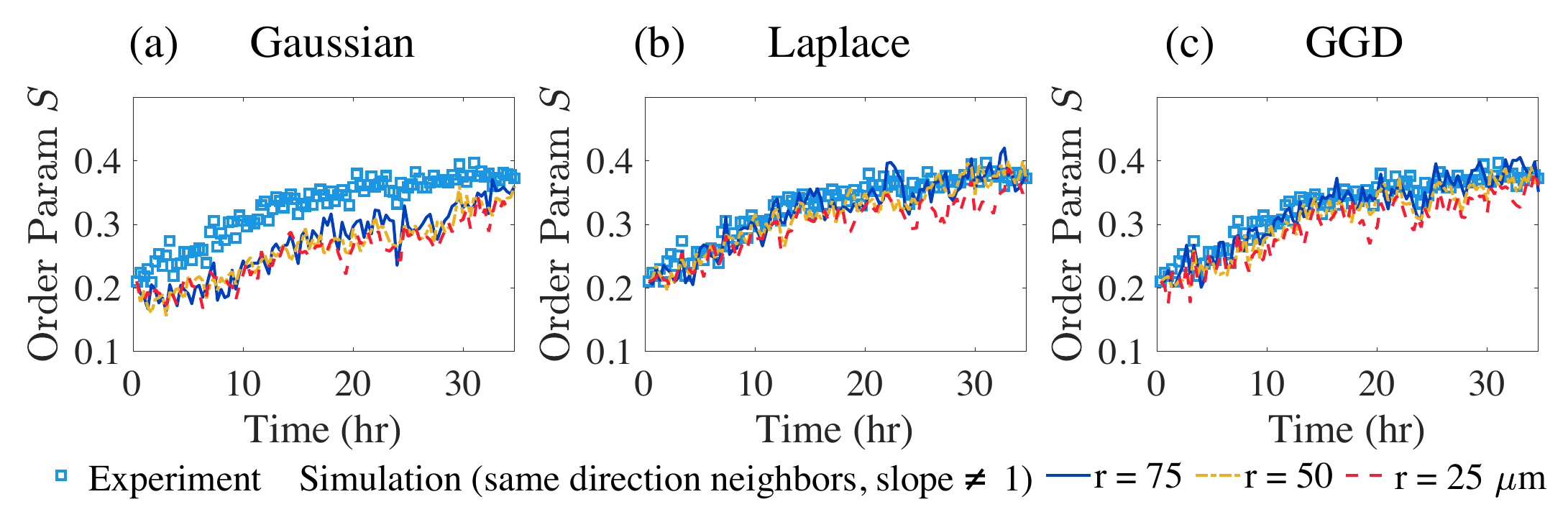}
   \caption{Reproducing orientational order parameters with different radii. The left, middle, and right panels compare simulations using the Gaussian, Laplace, and GGD
fluctuation with experimental observations. Open square symbols show the order parameter calculated from experimental data, dark blue solid line, yellow dash-dotted line, and red dashed line represent simulation results of different radii 75, 50, and 25 $\mu$m. } 
\label{fig:diff_r_sim_order_param_441}
\end{figure*}

The experimental setup in Fig. \ref{fig:diff_sim_order_param_sigma_109} has the same substrate material as the one presented in Fig. \ref{fig:diff_sim_order_param_sigma}. Except that at the start of imaging, cell density is higher. During the course of the experiment, the cell density  $\rho$ grows from $\rho$ = 500 to 650 mm$^{-2}$. 
The simulation model with selected features can accurately capture the progression of  orientational order parameter and velocity magnitude, shown in upper and lower panels  in Fig. \ref{fig:diff_sim_order_param_sigma_109}.
Models without all selected features cannot reproduce some of the properties. In particular, if Gaussian fluctuation is used with the other three features (blue curve in panel (a)), the model substantially underestimates the orientational order parameter, while the model with Laplace or GGD fluctuations, shown by blue curves in panel (b) and (c), respectively, reproduces the progression of these parameters reasonably well. Furthermore, all approaches incorporating estimated weights and considering neighbor ensembles of only cells traveling in the same direction capture the average absolute deviation of velocities $\tilde{\sigma}_x$ and $\tilde{\sigma}_y$  reasonably well.

On the other hand, results presented in Fig. \ref{fig:diff_sim_order_param_sigma_161} are derived from cell movements on an isotropic substrate, during this time, cell density changes from $\rho$ =  420 to 470 mm$^{-2}$. 
We have observed that, on the isotropic substrate, we reproduce order parameters that are around zero with any type of fluctuations  (Fig. \ref{fig:diff_sim_order_param_sigma_161}(a-c)), signaling the lack of order, and the velocity variance is also isotropic (Fig. \ref{fig:diff_sim_order_param_sigma_161}(d-f)).
It can be noted that when the cell moves on an isotropic substrate, choices of neighbor, fluctuation distributions, and weights become less important in fitting the order parameter (Fig. \ref{fig:diff_sim_order_param_sigma_161}(a-c)). 
Nonetheless, the absolute deviation$\tilde{\sigma}_x$ and $\tilde{\sigma}_y$ change over time due to proliferation. Approaches that utilize estimated weights and consider ensembles of neighboring cells traveling in the same direction effectively capture and accommodate these changes. 

\section*{Appendix E Additional results on simulation using different neighbor radii}

Here, we explore the effect of radii in reproducing the velocity order parameter for the entire monolayer. Given the cell width $\approx$ 20 $\mu m$ and length $\approx$ 100 $\mu m$ in projection, we explore length scales comparable to these dimensions. Below, we plot the order parameters reproduced by using different pair-wise distances: 25, 50, and 75 $\mu m$, in simulation (Fig. \ref{fig:diff_r_sim_order_param_441}). All simulations are reproduced  by including only the same direction neighbors and applying weights less than 1, as discussed before. Our results show that the variation in $r$ plays a much smaller role than changing the model for fitting the velocity fluctuations.

\end{document}